\documentclass[a4paper, 12pt]{article}
\usepackage[left = 1in, right = 1in, top = 1in, bottom = 1in]{geometry}
\usepackage{mathtools}
\usepackage{amsmath}
\usepackage{amssymb}
\usepackage{amsthm}
\usepackage{amsfonts}
\usepackage{physics}
\usepackage{bm}
\usepackage{nomencl}
\usepackage{graphicx}
\usepackage{xcolor}
\usepackage{pict2e}
\usepackage{mathrsfs}
\usepackage{caption}
\usepackage{lineno}
\usepackage{subfigure}
\usepackage{natbib}
\usepackage{color}
\usepackage{mathrsfs}
\usepackage[percent]{overpic}
\usepackage{hyperref}
\usepackage{appendix}
\usepackage{standalone}
\usepackage{float}
\usepackage{threeparttable}
\usepackage{authblk}

\def\R{\mathbb{R}}
\def\Rt{\mathbf{R}}
\def\r{\mathbf{r}}
\def\n{\mathbf{n}}
\def\m{\mathbf{m}}
\def\ea{\mathbf{e_{\alpha}}}
\def\eon{\mathbf{e_{1}}}
\def\eto{\mathbf{e_{2}}}

\def\e3{\mathbf{e_{3}}}
\def\di{\mathbf{d}_i}

\def\d3{\mathbf{d_{3}}}
\def\etal{\eta_\alpha}
\def\et3{\eta_3}
\def\rrho{\boldsymbol\rho}
\def\eps{\varepsilon}
\def\Q{\mathbf{Q}}
\def\E{\mathbf{E}}

\def\F{\mathbf{F}}

\def\U{\mathbf{U}}
\def\W{\mathbf{W}}
\def\H{\mathbf{H}}
\def\mH{\underline{H}}
\def\b{\boldsymbol\beta}
\def\k{\boldsymbol\kappa}
\def\vt{\underline{\text{v}}}
\def\kt{\underline{\text{k}}}
\def\nt{\underline{\text{n}}}
\def\mt{\underline{\text{m}}}
\def\v{\mathbf{v}}

\def\a{\boldsymbol a}
\def\w{\boldsymbol\omega}

\def\ps{\mathbf{\Psi}}
\def\psv{\boldsymbol\psi}

\def\ppi{\boldsymbol\Pi}
\def\pphi{\boldsymbol\Phi}
\def\Rot{\boldsymbol\Theta}

\title{{Twisting growth in rods with chiral material symmetries}}
\title{Buckling of chiral rods due to coupled axial and rotational growth}

\author[1]{ Satya Prakash Pradhan}
\author[2]{ Prashant Saxena\thanks{Corresponding author email: prashant.saxena@glasgow.ac.uk}}
\affil[1]{ \normalsize {Department of Mechanical \& Aerospace Engineering}\\
{Indian Institute of Technology Hyderabad, Sangareddy 502285, Telangana, India.}}
\affil[2]{ \normalsize {Glasgow Computational Engineering Centre, James Watt School of Engineering}\\
{University of Glasgow, Glasgow G12 8LT, UK.}}

\date{}

\begin{document}
% \vspace{-8ex}
  
\maketitle

% \addcontentsline{toc}{section}{Abstract}

\begin{abstract}
We present a growth model for special Cosserat rods that allows for induced rotation of cross-sections.
% Generalising elementary ideas of homogeneous growth, we arrive at a growth law with 
The growth law considers two controls, one for lengthwise growth and other for rotations. 
This is explored in greater detail for straight rods with helical and hemitropic material symmetries by introduction of a symmetry preserving growth to account for the microstructure. 
% growing rods that always remain straight. 
% A notion of symmetry preserving growth is introduced for rods with helical material symmetry to account for the induced rotation of cross-sections in such cases. 
The  example of a guided-guided rod possessing a chiral microstructure is considered to study its deformation due to growth. We show the occurrence of growth induced out-of-plane buckling in such rods. % corresponding to different bifurcation modes. 

\end{abstract}



% \tableofcontents

\section{Introduction}

%%\subsubsection*{Rod Theories and Applications}

Several theoretical models for elastic rods have been around for a while now. 
Starting from the Euler's elastica to Kirchhoff rods, a very rich literature is available including the general model developed by Green, Naghdi and their collaborators. 
The general rod theory proposed by Green and Naghdi subsumes classical theories like the Cosserat rod theory as special cases under appropriate constraints. 
A comprehensive description of different rod theories is provided by \cite{antman2005} and \cite{o2017modeling}.

Rod theories have been employed in many interesting  applications in the last few decades, for example in DNA biophysics \citep{manning1996continuum}, marine cables \citep{goyal2005nonlinear}, tendril perversion in plants \citep{goriely1998spontaneous,mcmillen2002tendril},  surgical filaments \citep{nuti2014modeling}, slender viscous jets \citep{arne2010numerical}, hair curls \textcolor{black}{\citep{miller2014shapes}} and carbon nanotubes \citep{chandraseker2009atomistic,kumar2011rod}. 

%  \cite{antman2005} and \cite{o2017modeling} give a comprehensive description of different rod theories.

% \subsection{Modelling biological growth using rod theories}

%Growing filamentary structures are ubiquitous in nature. Examples include bacterial fibres, fungi such as Phycomyces \citep{goriely2011spontaneous} and plant organs such tendrils, roots and stem \citep{wada2012hierarchical,wada2018twisting}. Some natural filaments tend to twist while growing axially \citep{wada2018twisting}. There are studies with helical growth models where straight axial growth is accompanied by rotation of cross-sections \citep{wada2012hierarchical,goriely2011spontaneous}. This type of twisting growth is the focus in our paper.

Growing filamentary structures are ubiquitous in nature. Plant organs such as tendrils, roots and stem tend to twist while growing axially \citep{wada2018twisting}. There are studies with helical growth models where straight axial growth is accompanied by rotation of cross-sections \citep{wada2012hierarchical,goriely2011spontaneous}. In this paper, we focus on this type of twisting growth, which can lead to non-planar configurations if the material of the rod exhibits some sort of twist-extension coupling. 

%Mathematical formulations of most growth problems in biology involve a multiplicative decomposition of the deformation gradient, a three dimensional version of which is present in \citep{rodriguez1994stress}. A one-dimensional version of the aforementioned multiplicative decomposition is present in \citep{moulton2013morphoelastic}. The recent study by \cite{moulton2020morphoelastic} gives a reduction of three-dimensional energy for a tubular structure to a one-dimensional equivalent via minimization in cross-sections and subsequent averaging; it further demonstrates the generation of intrinsic twists and curvatures caused by differential growth. A diverse account on biological growth is available in \citep{goriely2017mathematics}, containing both mathematical and bio-mechanical aspects. 

The standard multiplicative decomposition \citep{rodriguez1994stress} used to model biological growth has been specialised for one-dimensional growth by \cite{moulton2013morphoelastic}.
A recent study by \cite{moulton2020morphoelastic} gives the reduction of three-dimensional energy for a tubular structure to a one-dimensional equivalent via minimization in cross-sections and subsequent averaging; it further demonstrates the generation of intrinsic twists and curvatures due to differential growth. A diverse account on biological growth is available in \citep{goriely2017mathematics}, containing both mathematical and biomechanical aspects.

% Few notable works that model growing slender structures are mentioned here. 
Euler buckling of filaments evolving their shape under time varying loads has been considered by \cite{goldstein2006dynamic}. 
Works like 
\citep{mcmillen2002tendril} consider plant tendrils as Kirchhoff rods, straight in their initial states, which subsequently develop intrinsic curvatures in the grown equilibrium states. %, with appropriate boundary conditions. 
Another evolution law for intrinsic curvatures has been proposed by \cite{o2011evolution} with a focus on tip growth. 
% Some models 
\cite{guillon2012new} modelled tree growth by considering the branch to be a special Cosserat rod growing in both length and diameter. They modelled the reference, relaxed and current configuration of the growing rod with separate base curves and director fields.

% \subsection{Out-of-Plane Buckling}

 %, with different combinations of loading and boundary conditions.

%  , motivating us to look at twisting growth in such rods.
 
%  {\color{red} We also need some literature in this section for out of plane buckling in elastic rods without growth. Maybe check the references in \citep{healey2013bifurcation}.}

% \subsection{Material Symmetry and out-of-plane buckling}

\textcolor{black}{Several growing filaments in nature are known to have non planar configurations \citep{silverberg20123d,wada2012hierarchical,wada2018twisting}. Most existing works modelling one-dimensional growth in filaments stick to isotropic rods. Such models usually rely on differential growth \citep{moulton2020morphoelastic}, or presence of an external elastomeric matrix \citep{su2014buckling}, multi-rod composites \citep{lessinnes2017morphoelastic}, or phototropism \citep{moulton2020multiscale} to model the generation of curvature and torsion in non-planar configurations.
Growth in chiral rods can be another way to obtain such non-planar deformations; this has not been explored the literature. In this work, we show that growing chiral rods can buckle out of plane, simply with a boundary condition that arrests relative axial rotation at the ends.
}

%\textcolor{black}{In this work, we focus on growing chiral rods.}

There have been several attempts to understand material symmetries in rods. Different types of chiral material symmetries\textcolor{black}{-- such as hemitropy and helical symmetry-- in initially straight rods with uniform circular cross-section have been investigated in great detail by \cite{healey2002material}.} Other treatments of material symmetry in the context of rods include \citep{luo2000material,lauderdale2006transverse}, and to add which, \citep{lauderdale2007restrictions} draw a few parallel comparisons with some results by  \cite{healey2002material}.
In this manuscript, we follow the definitions and ideas of material symmetries introduced in \citep{healey2002material,healey2011rigorous}. 

Energy representations for helical symmetry and hemitropy have been derived in \citep{healey2002material}. Multi-fold helical symmetry is useful in modelling rods whose micro-structure mimics the symmetries of a rope made up of helices entwined together. Hemitropic rods possess the centre-line rotational symmetry of an isotropic rod but lack the reflection symmetries with respect to the longitudinal planes. Energy functions for rods with such chiral symmetries are typically characterized by coupled stretch, twist, shear and curvature terms. These couplings physically manifest as different types of non-traditional Poisson effects \citep{papadopoulos1999nonplanar}.  Moreover, the conventional quadratic energy densities associated with linear elasticity is incapable of distinguishing between different orders helical symmetries and hemitropy.  

Out-of-plane deformations are yet another feature of rods with such symmetries. Unshearable hemitropic rods can give rise to out-of-plane buckling when subjected to end displacements with fixed-fixed boundary condition, but on the other hand an axial load applied to a fixed-free rod always results in a planar solution \citep{healey2013bifurcation}. Similar bifurcation analysis has also been replicated for chiral rings with circular cross-sections under central loading \citep{hoang2019influence}. 
Both in-plane and out-of-plane buckling of isotropic rods embedded in elastomeric matrix have been examined by \cite{su2014buckling}, revealing that non-planar configurations are obtained whenever the matrix is stiff enough, compared to the bending stiffness of the rod.
%  Rods with chiral material symmetries are also known to buckle into non-planar shapes under appropriate loading and boundary conditions \citep{healey2013bifurcation}.
 Primary root growth of certain plants has been investigated by \cite{silverberg20123d}, drawing analogies from mechanical buckling of a metal filament embedded in a matrix comprising of two different gels whose interface is transverse to the filament.

% \subsection{Motivation}

% It is natural to expect rods with helical symmetry to twist while growing length-wise, however the exact relationship between the growth law and rod's microstructure is not well established. 
% However, this raises a fundamental question that how should such a chiral growth law be linked to the chiral constitutive law, or should they be linked at all. For instance, if the microstructure of a growing rod with helical symmetry gets modified during growth, its pitch is altered and the constitutive parameters controlling chirality must change accordingly. 

\textcolor{black}{The main focus of this work is to study} growth induced deformation in rods possessing chiral material symmetries \textcolor{black}{--- transverse hemitropy and dihedral helical symmetry}. \textcolor{black}{The growth law is also assumed to be chiral. Straight growth, where cross-sections do not rotate as they translate lengthwise, is not appropriate for modelling growth in rods with helical symmetry. If the chirality in material symmetry stems from some helical substructure associated with the rod's microstructure or from some sort of helical fibre-reinforcement, then simple translational growth can alter the pitch of the helix. This in turn may modify the chiral constitutive quantities associated with the material law. Moreover, straight translational growth without any rotation can lead to unwinding or over-winding, thus inducing additional stresses; stress-free growth in such cases requires the consideration of a coupled axial and rotational growth. Modelling the virtual configuration obtained from stress-free growth as a special Cosserat rod allows us to consider growth induced rotation of cross-sections.           
The exact relationship between the growth law and rod's microstructure is not well established. In this work,} we assume growth and constitutive laws to be independent in general. Additionally, for rods with helical symmetry we postulate the growth law to be \textit{symmetry preserving}, so that any imaginary helix associated with the microstructure remains unaltered as the rod grows. Such a growth problem depends only on the microstructural pitch and the constitutive laws, keeping aside the boundary conditions and other external factors. 

A rod constrained to grow (or decay) in a guided-guided environment is considered, with a chiral constitutive law that is applicable to both helical symmetry and transverse hemitropy. Out-of-plane buckling is observed to occur at certain growth (or atrophy) stages, corresponding to the bifurcation modes\textcolor{black}{.} %- - these deformed configurations are chiral in nature. 
We demonstrate that an exact reversal in chirality of these non-planar solutions requires us to mirror the chiral parameters in both growth and constitutive laws simultaneously. 
Comparisons are made for the end-to-end distance in the buckled configuration with that in the virtual state to see if the ends have come closer or moved apart, than what they would have been in the absence of the guides. 
% It may seem intuitive that an isotropic rod with a given growth law must have its end-to-end  distance between those obtained from any two rods with opposite material chiralities, growth law being the same. However we show that it is also possible to have otherwise.    
We also show that total growth induced extension in rod does not depend monotonically on the degree of chirality, that is, total extension in an isotropic rod need not lie between the total extension of rods with opposite material chirality.
% an isotropic rod with 

% We also explore all these aspects for the case of atrophy in the same problem set-up.

%  \subsection{Organisation}
This paper is organised as follows.
We begin with a theoretical background of material symmetries in the context of special Cosserat rods in Section \ref{sec:theoback}.
A twisting growth law with two control parameters is systematically derived using certain kinematic assumptions such as homogeneity in length-wise growth and relative rotation of cross-sections in Section \ref{sec:growthform}. 
In Section \ref{sec:guidedeg}, we solve the problem of growth induced out-of-plane bifurcation in a chiral rod with guided-guided boundary conditions to study the interplay between chiralities in growth and material laws.
We present our conclusions in Section \ref{sec: conclusions}.
%, essentially summarizing the ideas presented in \citep{healey2002material,healey2011rigorous}.
% We assume a stress-free, straight reference configuration that grows into the virtual configuration in the absence of environmental effects and boundary constraints, these external effects are then applied on the virtual state to get the current configuration. We consider separate multiplicative decomposition for the arc-length and the rotation field (Section \ref{sec:multdecmp}). 
% A twisting growth law is systematically derived using certain kinematic assumptions such as homogeneity in length wise growth and relative rotation of cross-sections in Section \ref{sec:homgrowthkine}. 
% The growth law obtained has two controlling parameters. 
% This is further extended to a case with special Cosserat reference configuration, not necessarily straight. Restricting ourselves to straight growth, where cross-sections are allowed to rotate only about the axis of the rod (Section \ref{sec:straightgrowth}), a helical growth law is arrived at. This is used further to examine the example of a rod with chiral material symmetry that is constrained to grow between coaxial guides. This guided-guided rod set-up is considered in Section \ref{sec:guidedeg} to study the interplay between chiralities in growth and material laws, while exploring the aspects of atrophy side by side.

\subsection{Notation}
% Most of the notations used in this paper are consistent with those used in \citep{healey2002material,healey2013bifurcation}. 
% Notation in this paper follows that of \citep{healey2002material}.
Throughout this text, the indices $i,j,k \in \{1,2,3\}$ and $\alpha,\beta\in \{1,2\}$, unless mentioned otherwise. 
We let $\{\eon,\eto,\e3\}$ to be a right-handed, fixed,  orthonormal basis for the Euclidean space $\mathbb{E}^3$. 
Boldface symbols are used to denote tensors, lowercase letters for first order e.g. $\mathbf{v}$  and uppercase letters for second order tensors e.g. $\mathbf{T}$. 
Underlined symbols such as $\underline{\text{v}}$ and $\underline{\text{T}}$ denote matrix representation of tensors with respect to a basis. %, not necessarily the fixed one. 

\section{Special Cosserat rod formulation} \label{sec:theoback}

\begin{figure}
\centering
\includegraphics[scale=0.72]{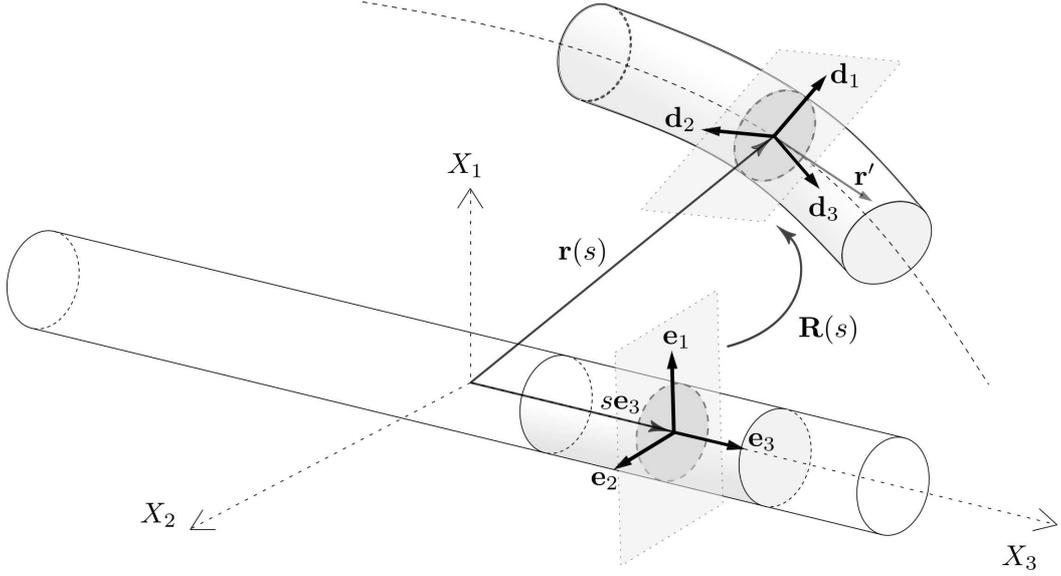}
\caption{Kinematics of a special Cosserat rod-- depicting the deformed centre-curve and the triad of orthonormal directors.}
\label{fig1}
\end{figure}

%% \subsection{Deformations in Rods}
Consider a straight rod of unit length in its stress-free reference configuration as shown in Figure \ref{fig1}.
Assumption of special Cosserat rod requires the transverse cross-sections to stay rigid during the deformation.
% throughout this text. 
% We stick to the special Cosserat rod theory which assumes the transverse cross-sections of the rod to be rigid. 
Let $s\in \big[-\frac{1}{2},\frac{1}{2}\big]$ denote a signed arc-length parameter of the centre-line in the reference configuration. % occupying $\{ s \e3 : -\frac{1}{2} \leq s \leq \frac{1}{2} \}$. 
Let $\r(s)$ define the centre-line of the deformed rod.
% , meaning that the point $s\e3$ in the reference configuration gets mapped to $\r(s)$ in the deformed configuration, $\r(s)$ is essentially a position vector with respect to some fixed origin (Figure \ref{fig1}). 
Let $\Rt(s)\in SO(3)$ be the rotation of transverse cross-sections in the reference configuration of the rod, mapping the fixed basis $\{\eon,\eto,\e3\}$ to a triad of orthonormal directors given by
\begin{align}
\mathbf{d}_i(s)=\Rt(s)\mathbf{e}_i.
\end{align} 

%
% Define the following 
The vector fields  
\begin{align}
\boldsymbol\nu:=\r'  ,\qquad  \k:=\text{axial}(\Rt'\Rt^T) \;
\end{align}
define convected coordinates $\boldsymbol\nu=\nu_i \di$ and $\k=\kappa_i \di$  with respect to the director frame field, along with the ordered triples $\vt:=(\nu_1,\nu_2,\nu_3)$ and  $\kt:=(\kappa_1,\kappa_2,\kappa_3)$.
The strains $\nu_\alpha$ correspond to shear, $\nu_3$ corresponds to stretch, $\kappa_\alpha$ correspond to curvatures, and $\kappa_3$ corresponds to twist. 

We further assume the rod to be hyperelastic with a differentiable energy density (per unit length) function $\Phi(\r',\Rt,\Rt',s)$.
Material objectivity  allows for a simpler version of energy function in terms of strains \citep{healey2002material}, given by
\begin{align}
\Phi=W(\vt,\kt,s),
\end{align} 
where $W$ is another differentiable scalar valued function. 

The internal force and moment on the transverse cross-section 
% originally at $s$ 
are denoted by $\n(s)=n_i \di$ and $\m(s) =m_i \di$, respectively,
% . We have the components $\n=n_i \di$ and $\m=m_i \di$, 
along with the corresponding triples $\nt :=(n_1,n_2,n_3)$ and $\mt :=(m_1,m_2,m_3)$. 
The components $n_\alpha$ are essentially the shear forces, $n_3$ is axial force, $m_\alpha$ are bending moments and $m_3$ is the torsional moment. 
These are related to the strain components as
\begin{align}
\nt=\frac{\partial W}{\partial \vt}
\quad , \quad 
\mt=\frac{\partial W}{\partial \kt}.
\end{align}

To prevent self penetration, we require
\begin{equation}
 \nu_3 =\r'\cdot\d3 >0 , \label{eq:nonpenetration}
\end{equation}
and the unshearability constraint is expressed as
\begin{equation}
 \nu_\alpha =\r'\cdot\mathbf{d}_\alpha =0. \label{eq:unshearability}
\end{equation}

% We make following standard assumptions. 
% \begin{align}
% \nu_3 &=\r'\cdot\d3 >0 .\label{eq:nonpenetration} \\
% \nu_\alpha &=\r'\cdot\mathbf{d}_\alpha =0. \label{eq:unshearability}
% \end{align}
% The assumption (\ref{eq:nonpenetration}) is known as the non-penetration condition and (\ref{eq:unshearability}) the unshearability constraint.

\subsection{Material symmetry in Rods}
%In this section, we present a brief overview of certain classes of material symmetry for special Cosserat rods \citep{healey2002material,healey2011rigorous}.
% review the mathematical definitions of a few chiral material symmetries in the context of special Cosserat rods \citep{healey2002material,healey2011rigorous}.
In this section, we present a brief overview of certain classes of material symmetry for special Cosserat rods, \textcolor{black}{as described by \cite{healey2002material,healey2011rigorous}}.
\subsubsection{Helical Symmetry}

\begin{figure}
\centering
\subfigure[\textcolor{black}{Rotating symmetry axis: each cross-section `$s$' in a rod with helical symmetry has a unique flip axis $\mathbf{e}_1^*(\frac{s}{\mathcal{M}})$ , a $180^\circ$ rotation about which gives the symmetry.}]{
\includegraphics[scale=0.72]{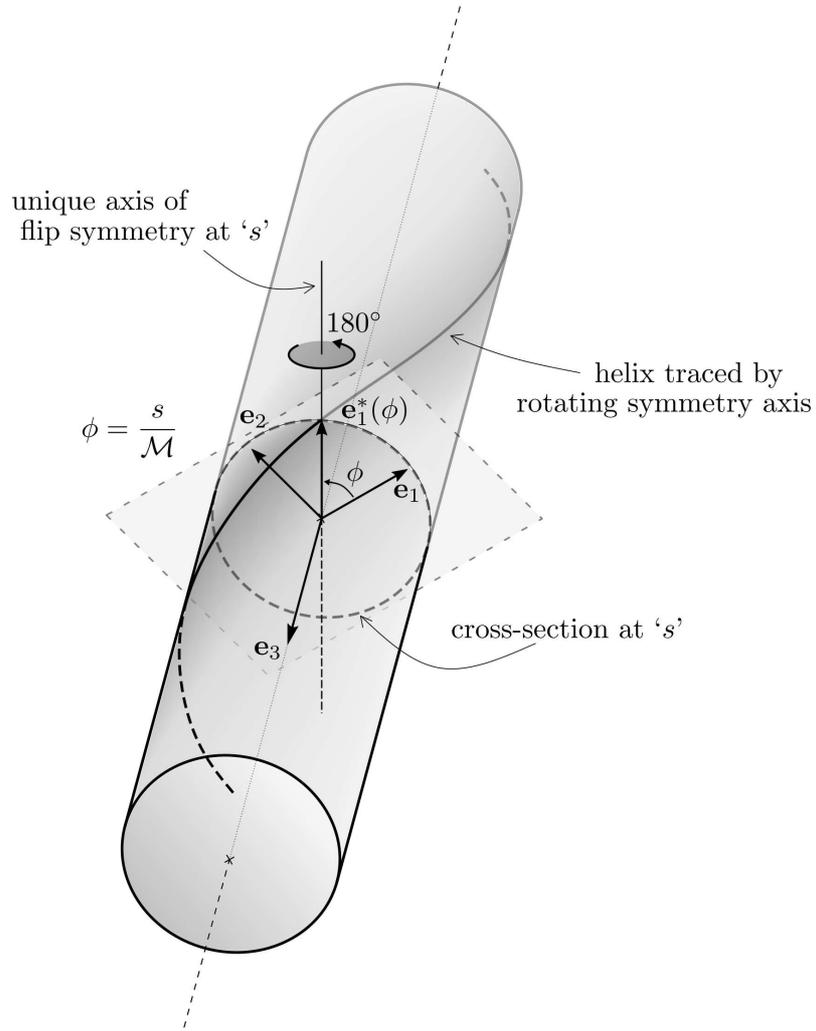}   \label{Flipaxisfig} }
\subfigure[\textcolor{black}{Heuristic depiction of a straight rod with 6-fold helical symmetry: it is analogous to six right handed helices entwined together, each with pitch $\mathcal{M}>0$.} ]{
\includegraphics[scale=0.72]{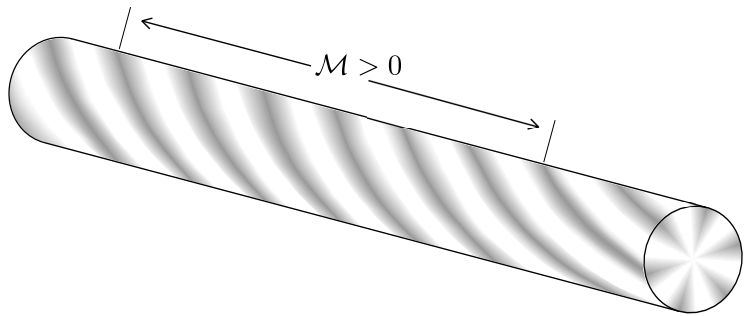}    \label{6fold}  }
\caption{\textcolor{black}{A depiction of symmetry and associated kinematic parameters in chiral rods \citep{healey2002material}.}}
\label{fig4}
\end{figure}

Consider a straight rod possessing helical material symmetry \textcolor{black}{\citep{healey2002material}.} 
% Assume $\mathcal{M}$ to be positive for right handed helices.
\textcolor{black}{ A unique flip axis (or symmetry axis)  is associated with every transverse cross-section that rotates as the section plane moves along the length of the rod (Figure \ref{Flipaxisfig}).}
% Every transverse cross-section \textcolor{black}{contains} a unique flip axis (or symmetry axis) which rotates as the section plane moves along the length of the rod \textcolor{black}{(Figure \ref{Flipaxisfig})}. 
A $180$-degree rotation (flip) about this axis renders the rod same as before. \textcolor{black}{We denote by $\mathcal{M}\neq 0$ its signed pitch, so defined that $\mathcal{M}>0$ for right-handed helices and $\abs{\mathcal{M}}$ is the least axial translation of the cross-section needed for the flip axis to complete a full rotation in $\eon-\eto$ plane (Figure \ref{6fold}).} % The flip axis completes a full rotation in the $\eon-\eto$ plane as the cross-section is moved by $\abs{\mathcal{M}}$, thus satisfying the standard definition of pitch.} 
%with a signed pitch $\mathcal{M}\neq 0$ with $\mathcal{M}>0$ for right-handed helices.

Unlike flips, reflections about a transverse plane do not result in a coincident helix, neither do the reflections through longitudinal planes. In fact, these reflections change the sign of $\mathcal{M}$, keeping its magnitude, the same.

%\begin{figure}[H]
%\centering
%\includegraphics[scale=0.72]{6folddihedralhelical}
%\caption{\textcolor{black}{Heuristic depiction of a straight rod with %6-fold helical symmetry: it is analogous to six right handed helices %entwined together, each with pitch $\mathcal{M}>0$.} }
%\label{6fold}
%\end{figure} 

\textcolor{black}{We introduce a rotating basis field
\begin{align}
\bigg\{\mathbf{e}^*_1\Big(\frac{s}{\mathcal{M}}\Big),\mathbf{e}^*_2\Big(\frac{s}{\mathcal{M}}\Big),\mathbf{e}^*_3\Big(\frac{s}{\mathcal{M}}\Big)=\mathbf{e}_3\bigg\} \label{eq:newbasis}
\end{align}
}
and a corresponding triad of director fields given by 
\begin{align}
\mathbf{e}^*_i(\phi) &=\textcolor{black}{\Rot}_\phi \mathbf{e}_i \;,\quad 0\leq \phi <2\pi  \\ 
\mathbf{d}^*_\alpha(s) &=\Rt(s)\mathbf{e}^*_\alpha\Big(\frac{s}{\mathcal{M}}\Big) ,
\end{align}
where $\textcolor{black}{\Rot}_\phi $ is a proper orthogonal tensor with matrix representation
%, whose matrix with respect to the fixed basis is
%[\Q_\phi]
\begin{align}
\textcolor{black}{\underline{\Theta}}_\phi =\begin{bmatrix}
\text{cos}\,\phi & -\text{sin}\,\phi & 0 \\ 
\text{sin}\,\phi & \text{cos}\,\phi & 0 \\ 
0 & 0 & 1 
\end{bmatrix}, \label{def:rot}
\end{align} 
in the fixed basis.

%\begin{figure}[H]
%\centering
%\includegraphics[scale=0.72]{flipaxisnewfinal}
%\caption{\textcolor{black}{Rotating symmetry axis: each cross-section %`$s$' in a rod with helical symmetry has a unique flip axis $\mathbf{e}%_1^*(\frac{s}{\mathcal{M}})$ , a $180^\circ$ rotation about which gives %the symmetry.}}
%\label{Flipaxisfig}
%\end{figure}

Assuming $\mathbf{e}^*_1(\phi)$ to be the rotating flip axis, %we
denote by $\H^\pi_\phi$ the flip about $\mathbf{e}^*_1(\phi)$, \textcolor{black}{so that
\begin{align}
\H^\pi_\phi \mathbf{e}^*_1(\phi)=\mathbf{e}^*_1(\phi) \;,
\qquad
\H^\pi_\phi \mathbf{e}^*_2(\phi)=-\mathbf{e}^*_2(\phi) \;,
\qquad
\H^\pi_\phi \mathbf{e}^*_3(\phi)=-\mathbf{e}_3.
\end{align}
} 
Material properties with respect to the symmetry axis $\mathbf{e}^*_1\big(\frac{s}{\mathcal{M}}\big)$ are assumed not to change as the cross-section `$s$' moves along the rod. This motivates the definition of a symmetry adapted energy function \citep{healey2002material} independent of $s$, given by
 \begin{align}
W(\vt,\kt,s)=\Phi=W^*(\vt^*,\kt^*),
\end{align}
where $\vt ^*=(\nu_1^*,\nu_2^*,\nu_3)$ and $\kt ^*=(\kappa_1^*,\kappa_2^*,\kappa_3) $ emerge from the change of coordinates
\begin{align}
\k =\kappa_\alpha^*\mathbf{d}_\alpha^*+\kappa_3\mathbf{d}_3 
\quad,\quad
\boldsymbol\nu =\nu_\alpha^*\mathbf{d}_\alpha^*+\nu_3\mathbf{d}_3 .
\end{align}
Helical symmetry is characterized by the following equation
\begin{align}
W^*(\nu_1^*,\nu_2^*,\nu_3,\kappa_1^*,\kappa_2^*,\kappa_3)=W^*(-\nu_1^*,\nu_2^*,\nu_3,-\kappa_1^*,\kappa_2^*,\kappa_3), \label{helicalsymm}
\end{align} 
in terms of the new energy function without `$s$' as an argument.

\subsubsection{n-fold Helical Symmetry}
Consider a rod with a symmetry analogous to $n\geq 2$ helices entwined together, such that each cross-section at $s$ has $n$ equally spaced flip axes. 
A $180$-degree rotation about each of these gives a symmetry \textcolor{black}{(Figure \ref{6fold})}. 
Such a rod is said to have a n-fold dihedral helical symmetry which is characterized by the condition 
\begin{align}
W^* \left(-\mH ^{\pi *}_{\frac{2\pi }{n}}\vt^*,-\mH ^{\pi *}_{\frac{2\pi }{n}}\kt^* \right)=W^*(\vt^*,\kt^*) , \label{nhelicalsymm}
\end{align}
in addition to \eqref{helicalsymm}, where $\mH ^{\pi *}_{\frac{2\pi }{n}}$ is the matrix of \textcolor{black}{$\H ^{\pi }_{\frac{2\pi }{n}}$} with respect to the rotating basis \eqref{eq:newbasis}.
\begin{align}
\mH ^{\pi *}_{\frac{2\pi }{n}}=\begin{bmatrix}
\text{cos}(\frac{2\pi }{n}) & \text{sin}(\frac{2\pi }{n}) & 0 \\[7.5pt] 
\text{sin}(\frac{2\pi}{n}) & -\text{cos}(\frac{2\pi }{n}) & 0 \\[7.5pt] 
0 & 0 & -1 
\end{bmatrix}.
\end{align}

\subsubsection{Continuous Helical Symmetry}
For $n \gg 1$, a straight rod with $n$-fold dihedral helical symmetry approaches to what is called continuous helical symmetry. In this type of symmetry 
all vectors of the cross section act as symmetry axis, or equivalently any fixed flip axis, say $\eon$ acts as a symmetry axis for all cross sections. 
Continuous helical symmetry can be characterized by 
\begin{align}
W(-\mH^\pi_\phi\vt,-\mH^\pi_\phi\kt)=
W(\underline{\text{v}}, \underline{\text{k}}),\quad \forall\, \phi\in [\,0,\pi).
\end{align}

\subsubsection{Transverse hemitropy and isotropy}
Let $\E$ denote \textcolor{black}{the} reflection with matrix 
\begin{align}
\underline{E}=
\begin{bmatrix}
1 & 0& 0 \\ 
0 & -1 & 0 \\ 
0 & 0 & 1 
\end{bmatrix},
\end{align}
\textcolor{black}{written in the fixed basis.} 
A homogeneous hyperelastic straight rod with energy function $W(\vt,\kt)$ is transversely hemitropic if
\begin{align}
W(\underline{\textcolor{black}{\Theta}}_\phi\vt,\underline{\textcolor{black}{\Theta}}_\phi\kt)=W(\vt,\kt)\quad\forall\;\phi\in [\,0,2\pi) , \label{eq:defhemitropy}
\end{align}
and flip-symmetric if  
\begin{align}
W(\underline{E}\vt,\underline{E}\kt)=W(\vt,\kt) . \label{eq:flipsymmcond}
\end{align}
Note that flip-symmetry does not belong to the class of transverse symmetry, defined by \cite{healey2002material}.
A straight rod is transversely isotropic if in addition to \eqref{eq:defhemitropy}, it also satisfies
\begin{align}
W(\vt,\kt)=W(\underline{E}\vt,-\underline{E}\kt).
\end{align}
 Flip-symmetric hemitropy is equivalent to continuous helical symmetry \citep{healey2002material}. Another way to obtain flip-symmetric hemitropy, is to consider a rod with helical symmetry and take the limit $\mathcal{M} \to 0$ \citep{healey2011rigorous}.

\subsection{Energy function}
The energy density per unit length of unshearable hemitropic rods can be expressed as \citep{antman2005,healey2002material}
\begin{align}
W=\Upsilon( \kappa_\alpha\kappa_\alpha , \nu_3 , \kappa_3)  , \label{eq:energyunshearable}
\end{align}
where $\Upsilon$ is a scalar valued function. 
This representation is also valid for flip-symmetry.
For calculations in this paper, we adopt a model considered by \cite{papadopoulos1999nonplanar,healey2013bifurcation} defined as
% 
%  Consider the function
\begin{align}
\Upsilon= \frac{1}{2}\Big[\textcolor{black}{\Psi}(\nu_3)+2A(\nu_3-1)\kappa_3+B\kappa_3^2+C\kappa_\alpha\kappa_\alpha\Big], \label{densnearquad}
\end{align}
where $\textcolor{black}{\Psi}:(0,\infty)\rightarrow \R$ is a function such that $g:=\frac{1}{2}\textcolor{black}{\Psi}'$ obeys $g(\nu_3)\rightarrow-\infty$ as $\nu_3 \rightarrow 0$.
The function $g(\cdot)$ allows us to modify the axial force response of the model, and it must satisfy $g(1)=0$. The constant $C$ corresponds to bending stiffness, $B-\displaystyle\frac{A^2}{g'(1)}$ is equivalent to torsional rigidity and $g'(1)-\displaystyle\frac{A^2}{B}$ to axial stiffness, where $A$ is the degree of hemitropy. We assume $B>0$, $C>0$ and $Bg'(\nu_3)>A^2$ for all $\nu_3$  to ensure convexity. This in turn implies that $g(\cdot)$ should be monotonic and hence invertible.
For example, a response function  satisfying all our criteria can be chosen as \citep{papadopoulos1999nonplanar}
\begin{align}
g(\nu_3)= F\, \text{ln}(\nu_3)+ \frac{A^2}{B}(\nu_3-1), \label{eq:axresfun}
\end{align} 
where $F>0$ is a constant.
This energy allows for infinite compressive axial force $n_{3}\rightarrow-\infty$ whenever an unrealistically extreme strain $\nu_3 \rightarrow 0$ is present.

As demonstrated in \citep{healey2002material}, quadratic energy functions are incapable of distinguishing between  different types of $n$-fold helical symmetry ($n\geq 3$) and hemitropy.
On similar lines, the energy function \eqref{densnearquad} can be shown to be applicable to $n$-fold helical symmetry.
 
\section{Growth Formulation} \label{sec:growthform}

Growth in elastic bodies is typically modelled by introduction of a multiplicative decomposition of the deformation gradient into pure growth and pure elastic deformation parts \citep{rodriguez1994stress,ambrosi2011perspectives}.
This decomposition assumes a virtual stress-free incompatible configuration.
% For 1-D structures
% Biological growth in elastic bodies is often split into the following hypothetical steps to aid in its mathematical treatment. First, the body is virtually cut into pieces, so that no residual stress is present in the individual pieces, each of which are then let grow separately without developing any residual stress or incompatibility in this virtual configuration. These individually grown pieces can then be forcibly put together and allowed to relax to give the current deformed configuration. This process is mathematically represented by a multiplicative decomposition of the deformation gradient \citep{rodriguez1994stress,ambrosi2011perspectives}. 
For one-dimensional structures where growth manifests as increase in overall length, first the stress-free rod isolated from its environment and boundary conditions can be allowed to grow free into a virtual state, and then the boundary and environmental factors can be forcibly imposed  \citep{goriely2017mathematics,o2013growth}.

\textcolor{black}{But one-dimensional growth models, where cross-sections simply translate during free growth are not suitable for several classes of chiral rods. 
Chiral rods usually have a physical winding bias intrinsic to the microstructure \citep{healey2002material}. Length-wise growth with no cross-sectional rotation can modify this microstructure. For example, a rod with helical symmetry made to grow axially will have to change its inherent pitch if the cross-sections are not allowed to rotate during growth; and as a result the constitutive parameters controlling material chirality must change accordingly. In such examples, to be able to look at growth that does not alter the microstructure, or restricts the microstructure to modify itself in a particular manner, it is essential that we look at rod growth in a more general setup. }

\textcolor{black}{
Similarly in chiral rods where material symmetry arises from fibre-reinforcement \citep{shirani2020cosserat}, rod growth is a result of individual fibre growth and it is the growth pattern of these fibres that dictates whether the rod's cross-sections must rotate, as they get translated during growth. Consider a rod that is composed of fibres twisted helically in the unstressed reference state; if the cross-sections are not allowed to rotate during growth, it would have an unwinding or over-winding effect on the fibres thus generating stresses. In such cases for growth to take place without the generation of any stress, the cross-sections must rotate.}

\textcolor{black}{
 This is why we choose to individually treat all three configurations (reference, virtual and current) as special Cosserat rods, and then analyse the relative rotations. }

\subsection{General Framework for Growing Rods}\label{sec:multdecmp}

\begin{figure}
\centering
\includegraphics[scale=0.88]{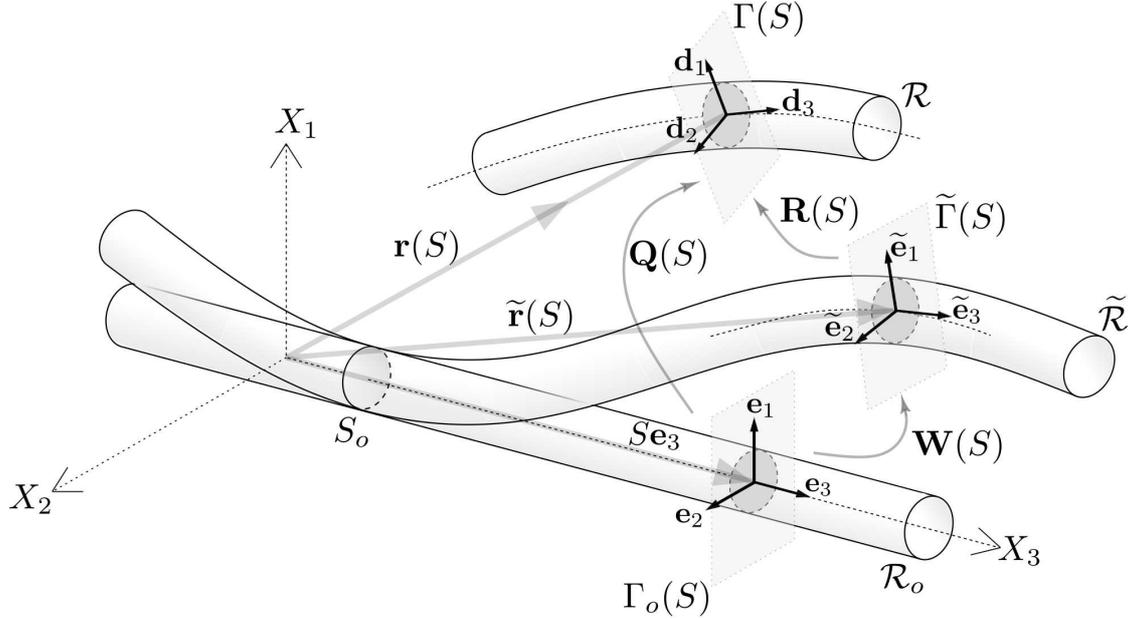}
\caption{Kinematics of an initially straight rod growing from origin $S_o$, depicting the configurations-- reference $\mathcal{R}_o$, virtual $\widetilde{\mathcal{R}}$ and  current $\mathcal{R}$; along with the multiplicative decomposition $\Q=\Rt\W$.}
\label{fig2}
\end{figure}

Let $\mathcal{R}_o$ denote the initial stress-free reference configuration of the rod, occupying $\{ S \e3 : -\frac{1}{2} \leq S \leq \frac{1}{2} \}$ and denote by $S$ a signed arc length parameter of the centre-line in $\mathcal{R}_o$.
Let $\widetilde{\r}(S)$ be the curve taken by the centre-line in the virtual grown configuration $\widetilde{\mathcal{R}}$, such that the point $S\e3$ in $\mathcal{R}_o$ gets mapped to $\widetilde{\r}(S)$ in $\widetilde{\mathcal{R}}$ (Figure \ref{fig2}). 
The virtual configuration is assumed to be stress-free. 
We define a signed  arc-length $s(S)$ in $\widetilde{\mathcal{R}}$ by
\begin{align}
s(S):=\displaystyle\int_0^{S}\norm{\widetilde{\r}\,'(\tau)}d\tau \textcolor{black}{,}
\end{align}
\textcolor{black}{where $\norm{\cdot}$ denotes the Euclidean vector norm.}
We denote the transverse cross-section at $S$ in $\mathcal{R}_o$ by $\Gamma_o(S)$ and let it get mapped to $\widetilde{\Gamma}(S)$ in the virtual configuration $\widetilde{\mathcal{R}}$.
Define $\W(S)\in SO(3)$ to be the rotation of $\widetilde{\Gamma}(S)$ with respect to $\Gamma_o(S)$, and let it map the fixed basis $\{\eon,\eto,\e3\}$ to a virtual director field given by
\begin{align}
\widetilde{\mathbf{e}}_i(S)=\W(S)\mathbf{e}_i.
\end{align}   
When the boundary conditions and environmental factors are imposed, let the centre-line take the curve $\r(S)$ in the current configuration $\mathcal{R}$, and the cross-section $\widetilde{\Gamma}(S)$ in $\widetilde{\mathcal{R}}$ be mapped to $\Gamma(S)$ in $\mathcal{R}$. Define $\Rt(S)\in SO(3)$ to be the rotation of $\Gamma(S)$ with respect to $\widetilde{\Gamma}(S)$ and $\Q(S)\in SO(3)$ to be the \textcolor{black}{net} rotation of $\Gamma(S)$ with respect to $\Gamma_o(S)$, so that
\begin{align}
\Q(S)=\Rt(S)\W(S) .
\end{align}
 The virtual director field is transformed into another director field in the current configuration given by 
\begin{align}
\mathbf{d}_i(S)=\Rt(S)\widetilde{\mathbf{e}}_i(S)=\Q(S)\mathbf{e}_i.
\end{align}

All the maps we have introduced are assumed to be smooth for the sake of convenience. Analogous to $\r:\big[-\frac{1}{2},\frac{1}{2}\big]\rightarrow\mathbb{E}^3$, we define another map $\widehat{\r}:\big[s(-\frac{1}{2}),s(\frac{1}{2})\big]\rightarrow\mathbb{E}^3$ to denote the same curve via \textcolor{black}{the reparametrization},   
\begin{align}
\r(S)=\big(\widehat{\r}\circ s\big) (S).
\end{align}
This implies
\begin{align}
\r'(S)=\norm{\widetilde{\r}\,'(S)}\textcolor{black}{\frac{\partial\widehat{\r}}{\partial s},}
\end{align}
\textcolor{black}{
where $(\cdot)':=\displaystyle\frac{\partial}{\partial S}\big(\cdot\big)$;
taking the magnitude of which gives the one-dimensional multiplicative decomposition
\begin{align}
 \norm{\frac{\partial \r}{\partial S}}=\norm{\frac{\partial \widetilde{\r}}{\partial S}}\norm{ \frac{\partial \widehat{\r}}{\partial s} }.
\end{align}
}
Similarly, define $\widehat{\Rt}:\big[s(-\frac{1}{2}),s(\frac{1}{2})\big]\rightarrow SO(3)$ by 
\begin{align}
\Rt(S)=\big(\widehat{\Rt}\circ s\big) (S).
\end{align}
We assume the transverse cross-sections to remain orthogonal to centre-line in both virtual and current configurations, hence the conditions 
\begin{align}
\widetilde{\r}\,'\cdot\widetilde{\mathbf{e}}_\alpha &=0 \label{eq:growthtangency} \\
\text{and }  
\r'\cdot\mathbf{d}_\alpha &=0, \label{eq:newunshearability}
\end{align}
where \eqref{eq:newunshearability} is equivalent to the unshearability constraint \eqref{eq:unshearability}.
The symbols and notations introduced in this section are pictorially represented in Figure \ref{fig2}.

\subsubsection{Homogeneous growth kinematics}   \label{sec:homgrowthkine}
We consider the growth to be homogeneous throughout the rod. 
This assumption leads to the following constraints:
% We restrict the growth law according to the following assumptions.
% Now we make few kinematic assumptions on rod growth. Let $h\in\R$ be such that $0<|h|<1$.
\begin{itemize}
  \item The length-wise growth parameter denoted by 
  $ \gamma:=||\widetilde{\r}\,'(S)||$ is a constant, that is, it is independent of $S$.
%   \item 
%   As the rod grows, points  $`S$' and $`S+h$' on the center-line of reference configuration move apart to a distance $s(S+h)-s(S)$, along the arc-length of the virtual configuration, which is assumed to be independent of $S$. This implies,
% \begin{align}
% \gamma:=||\widetilde{\r}\,'(S)||  \text{ is thus a constant for all } S. 
% \label{}
% \end{align}

\item 
Let $h\in\R$ be such that $0<|h|<1$.
Consider the relative rotation of cross-section $\widetilde{\Gamma}(S+h)$ with respect to $\widetilde{\Gamma}(S)$.
\begin{align}
 \widetilde{\mathbf{e}}_i(S+h)=\W(S+h)\W(S)^{-1}\,\widetilde{\mathbf{e}}_i(S). \label{unirelrotgr}
\end{align}
For all permissible $h$, the relative rotation $\W(S+h)\W(S)^{-1} $ is assumed to be independent of $S$, and hence can be denoted as a function of $h$ only.
\begin{align}
\W(S+h)\W(S)^{-1}=:\ppi(h).\label{eq:rotahomo} 
\end{align}
\textcolor{black}{
This gives us the decomposition
\begin{align}
\W(S+h)=\ppi(h)\W(S).
\end{align}
Choosing $\hbar \neq 0$ such that all the tensor fields appearing in the following calculation make sense, we have
\begin{subequations} 
\begin{align}
\Big\{\ppi(h+\hbar)-\ppi(h)\Big\}\W(S)&=\W(S+h+\hbar)-\W(S+h)\\
&=\ppi(h)\Big\{\W(S+\hbar)-\W(S)\Big\}.
\end{align}
\end{subequations}
Dividing by $\hbar$ and taking the limit $\hbar\rightarrow 0$ yields
\begin{align}
\frac{\partial\ppi(h)}{\partial h}\W(S)=\ppi(h)\frac{\partial\W(S)}{\partial S}. \label{eq:depres}
\end{align}
We define the following tensor fields for our convenience,
\begin{align}
\mathbf{\Lambda}(S):=\W(S)^T\,\frac{\partial \W(S)}{\partial S} ,
\quad \text{and} \quad 
\mathbf{\Omega}(S):=\frac{\partial \W(S)}{\partial S}\W(S)^T .
\label{eqtensdiff}
\end{align}
Now equation \eqref{eq:depres} implies 
\begin{align}
\ppi(h)^T\frac{\partial\ppi(h)}{\partial h}=\mathbf{\Omega}(S).\label{eq:constensor1}
\end{align}
Since $h$ and $S$ can be chosen arbitrarily, independent of each other, we conclude that $\mathbf{\Omega}(S)$ is constant. 
Another way to interpret \eqref{eq:rotahomo} is to set
\begin{align}
\frac{\partial}{\partial S}\Big\{\W(S+h)\W(S)^{-1}\Big\}=\textbf{O}\,, \label{diffrothom}
\end{align}
We expand the derivative to get
\begin{subequations}
\begin{align}
\frac{\partial}{\partial S}\Big\{\W(S+h)\W(S)^{-1}\Big\}
&=\frac{\partial}{\partial S}\big\{\W(S+h)\big\}\W(S)^T-\W(S+h)\W(S)^T\frac{\partial\W(S)}{\partial S}\W(S)^T\\
&=\W(S+h)\Big\{\mathbf{\Lambda}(S+h)-\mathbf{\Lambda}(S)\Big\}\W(S)^T.
\end{align}
\end{subequations}
This implies $\mathbf{\Lambda}(S+h)=\mathbf{\Lambda}(S)$ for all choices of $S$ and $h$, chosen independent of each other, which means $\mathbf{\Lambda}(S)$ is constant. \\
Moreover, these constant tensors $\mathbf{\Lambda}$ and $\mathbf{\Omega}$ are skew-symmetric (Appendix \ref{Apx:A}). }

% \vspace{10pt}

% \vspace{5pt}

\item
We fix a point on the centre-line which gets mapped to itself under the growth transformation, along with its corresponding cross-section. Thus, we assume the existence of a point $S_o \in \big[-\frac{1}{2},\frac{1}{2}\big]$ satisfying
\begin{align}
\widetilde{\r}(S_o)=S_o \e3 
\qquad\text{and}\qquad
\W(S_o)=\mathbf{I}.
\end{align}
This can also be interpreted as if the rod is allowed to grow while being held at $S_o$ (origin of growth). It is held in such a way that no incompatibility or stress is caused due to growth. % (Figure \ref{fig2}).
Define vectors $\a:=\text{axial}\big(\mathbf{\Lambda}\big)$ and $\w:=\text{axial}\big(\mathbf{\Omega}\big)$, these are actually constant vectors and can be related by 
\begin{align}
\w=\W(S)\a \,.
\end{align} 
Since this is also satisfied for the specific point $S=S_o$, we imply $\a=\w$ and $\mathbf{\Lambda}=\mathbf{\Omega}$.
This also means that \textcolor{black}{$\text{axis}\big(\W(S)\big)=\a$} for all $S$. 
\textcolor{black}{
Thus one can solve the system
\begin{align}
\W^T \frac{\partial \W}{\partial S}=\mathbf{ \Lambda } \quad \text{with} \quad
\W(S_o)=\mathbf{I},  \label{Wdiffeqn}
\end{align}
for  $\W$ to obtain 
\begin{align}
\W(S)=e^{(S-S_o) \mathbf{ \Lambda } }, \label{growth}
\end{align}
}
where tensor exponential is defined by the usual series definition. The mathematical details for derivations in this section are provided in Appendix \ref{Apx:A}.
\end{itemize}

\subsubsection{Extension to a general growing curve}
Consider a general scenario where the initial configuration $\mathcal{R}_o$ is a special Cosserat rod. Let $\bar{\r}:\big[-\frac{1}{2},\frac{1}{2}\big]\rightarrow\mathbb{E}^3$ be its centre-curve, where $\bar{\r}(S)$ is arc-length parametrized. Let $\bar{\W}(S)\in SO(3)$ denote the orientation of $\Gamma_o(S)$ with respect to the fixed basis, mapping those to an orthonormal director field $\bar{\mathbf{e}}_i(S):=\bar{\W}(S)\mathbf{e}_i$ associated with initial configuration. 
\textcolor{black}{Since $\W(S)\in SO(3)$ maps $\Gamma_o(S)$ to $\widetilde{\Gamma}(S)$, the virtual director field should be given by
$\widetilde{\mathbf{e}}_i(S)=\W(S)\bar{\W}(S)\mathbf{e}_i$; so that}
 equation \eqref{unirelrotgr} is modified as
 \begin{align}
 \widetilde{\mathbf{e}}_i(S+h)=\W(S+h)\bar{\W}(S+h)\bar{\W}(S)^{-1}\W(S)^{-1} \widetilde{\mathbf{e}}_i(S) .
  \end{align} 
% Adhering to rest of the growth related notations introduced in the previous section, the condition (\ref{unilengr}) for length-wise uniformity in growth still holds. 

 \textcolor{black}{Homogeneous growth law still requires $\gamma$ to be constant.}
The tensor $\W(S+h)\W(S)^{-1} $ is again independent of $S$.
%  but the tensor we require to be independent of $S$ is still $\W(S+h)\W(S)^{-1} $.
%  To see this imagine a sub-rod segmented between the cross-sections $\Gamma_o(S)$ and $\Gamma_o(S+h)$. In the grown configuration $\Gamma_o(S)$ rotates by $\W(S)$ to become $\widetilde{\Gamma}(S)$. But if the same rotation is applied to the whole sub-rod, an additional rotation of $\W(S+h)\W(S)^{-1} $ would still be required at the $S+h$ end for it to become $\widetilde{\Gamma}(S+h)$. Thus homogeneity requires $\W(S+h)\W(S)^{-1} $ to be dependent on $h$ only.
 In addition, the rod is assumed to be held at $S_o \in \big[-\frac{1}{2},\frac{1}{2}\big]$ while growing, so that we have
\begin{align}
\widetilde{\r}(S_o)=\bar{\r}(S_o) \qquad\text{and}\qquad
\W(S_o)=\mathbf{I}.
\end{align}
This assumption along with the kind of homogeneity used in induced rotations gives such a $\W(S)$ that makes all the cross-sections rotate about the particular axis $\a$. Moreover, the solution is given by \eqref{growth} which in turn implies
\begin{align}
\widetilde{\mathbf{e}}_i(S)= e^{(S-S_o)\mathbf{ \Lambda }}\bar{\W}(S)\mathbf{e}_i, \label{eq:gengrowth}
\end{align}
In fact, the constant vector $\w=\a$ can be treated as the growth parameter controlling relative rotation of cross-sections while $\gamma$ controls the length-wise growth as in the former case. Whenever the centre-curves are normal to the cross-sections, throughout $\mathcal{R}_o$ and $\widetilde{\mathcal{R}}$, we deduce
\begin{align}
\widetilde{\r}(S)=\bar{\r}(S_o)+\gamma\!\int_{S_o}^S \!e^{(\tau-S_o)\mathbf{ \Lambda }} \bar{\r}\,'(\tau)\,d\tau \,.
\end{align}
We emphasise that \eqref{eq:gengrowth} and \eqref{growth} do not assume the respective centre-curves to be normal to the cross-sections neither in  $\mathcal{R}_o$ nor in $\widetilde{\mathcal{R}}$.

\subsection{Growth in straight rods} \label{sec:straightgrowth}
Consider a straight rod with flip-symmetric hemitropy in its reference configuration. 
% All notions of material symmetry in a rod \citep{healey2002material} assume it to be straight in undeformed state. 
% To be able to follow the definitions of symmetry introduced in \citep{healey2002material}, we assume free growth to so restrict the rod that it always remains straight. 
A straight virtual configuration  condenses to 
\begin{align}
\widetilde{\r}(S)=\{S_o+ \gamma(S-S_o) \}\e3 ,
\end{align}
which with the aid of \eqref{eq:growthtangency} results in
\begin{align}
\W(S)\e3=\e3\quad \forall S \in \Big[-\frac{1}{2},\frac{1}{2}\Big].
\end{align}
This indicates that $\w$ is along $\e3$. We introduce another growth parameter $\omega$ defined by 
\begin{align}
\w=\omega\e3 ,
\end{align} 
so that its corresponding skew tensor is
\begin{align}
\mathbf{\Omega} =\omega\mathbf{A}, \quad \text{with} \quad
\mathbf{A} =\eto \otimes \eon- \eon \otimes \eto.
\end{align}
Since rotation tensor can also be expressed as
\begin{align}
\textcolor{black}{\Rot}_\phi=e^{\phi\mathbf{A}},
\end{align}
we get
\begin{align}
\W(S)=\textcolor{black}{\Rot}_{(S-S_o)\omega}.
\end{align}
 
The parameters $\gamma$ and $\omega$ capture all the necessary information regarding growth. \textcolor{black}{It is} evident that $\gamma>1$ reflects growth while $\gamma<1$ denotes atrophy. Similarly, $\omega$ and $-\omega$ signify two opposite cross-sectional rotations caused by growth while $\omega=0$ indicates no growth induced rotation.

\textcolor{black}{This type of growth is helical in nature. Consider any line in the bulk of the rod parallel to its axis, but not the axis itself. As the rod grows this line transforms into a helix of pitch $\displaystyle\frac{\gamma}{\omega}$. Similarly any helix in the initial configuration transforms into another coaxial helix, not necessarily with the same pitch. This is a reflection of the fact that the Darboux vector of a helix is a constant vector aligned along its axis.}  

\subsubsection{Growth law}
% The growth law adopted demands 
% A growth law for rods requires 
The growth law adopted here considers rotation of cross sections with respect to each other in the due course of growth.
% Consider any line in the bulk of the rod parallel to its axis, but not the axis itself. As the rod grows this line transforms into a helix of pitch $\displaystyle\frac{\gamma}{\omega}$. Similarly any helix transforms into a helix, not necessarily of same pitch. To see this 
Consider a rotating basis field $\{\mathbf{e}^*_1(S),\,\mathbf{e}^*_2(S),\,\mathbf{e}^*_3(S)=\e3 \}$ given by 
\begin{align}
\mathbf{e}^*_i(S)=\textcolor{black}{\Rot}_{\frac{S}{\mathcal{M}}}\mathbf{e}_i,
\end{align}
representing a helix embedded in the initial configuration of a rod. As the rod grows this transforms into $\W(S)\mathbf{e}^*_i(S)$ in the virtual configuration. Let us denote this by a basis field $\{\mathbf{f}^*_1(s),\,\mathbf{f}^*_2(s),\,\mathbf{f}^*_3(s) \}$ defined on the virtual arc-length parameter by
\begin{align}
\W(S)\mathbf{e}^*_i(S)=: \big(\mathbf{f}^*_i\circ s\big) (S).
\end{align}    
This is equivalent to 
\begin{align}
\mathbf{f}^*_i(s)=\textcolor{black}{\Rot}_{\frac{s}{\gamma\mathcal{M}}+(\frac{s}{\gamma}-S_o)\omega}\mathbf{e}_i .
\end{align}
Let $h \neq 0$ be such that $\mathbf{e}^*_i(S+h)$ and $\mathbf{f}^*_i(s+h)$ are well defined, then we obtain
\begin{align}
\mathbf{e}^*_i(S+h)&=\textcolor{black}{\Rot}_{\frac{h}{\mathcal{M}}}\mathbf{e}^*_i(S) ,\\
\mathbf{f}^*_i(s+h)&=\textcolor{black}{\Rot}_{\frac{h}{\gamma}(\frac{1}{\mathcal{M}}+\omega )}\mathbf{f}^*_i(s) .
\end{align}
This shows that our chosen growth map transforms the initial  helix with pitch $\mathcal{M}$ into another helix with pitch, say $\mu$, which can be expressed as 
\begin{align}
\mu=\frac{\gamma \mathcal{M}}{1+\omega \mathcal{M}}.\label{eq:pitchevolve}
\end{align} 
This motivates us to define a \textit{symmetry preserving} growth law for rods possessing helical symmetry.

\paragraph{Rods with helical symmetry}

Consider a rod which due to its microstructure possesses simple helical symmetry or $n$-fold helical symmetry. Let $\mathcal{M}$ be the pitch associated with its microstructure. Once growth parameters $\gamma$ and $\omega$ are known, \eqref{eq:pitchevolve} serves as an evolution law for the pitch of its microstructure.

 We introduce the idea of \textit{symmetry preserving} growth -- wherein the growth map fixes all helices with pitch same as that of the microstructure ($\mu = \mathcal{M}$). Thus, for rods with a pitch associated with their microstructure we have the following helical growth law
 \begin{align}
\gamma=1+\omega \mathcal{M}, \label{eq:pitchfix}
\end{align}
where $\gamma$ is the only growth parameter and $\mathcal{M}$ comes from the material symmetry. For rods having helical symmetry, this assumption of \textit{symmetry preserving} growth provides a rationale for relative rotation of cross-sections during growth. 

\paragraph{Hemitropic rods}
Although there are different versions \citep{healey2002material,healey2011rigorous} of how helical symmetry can be used to arrive at hemitropy, there is no pitch directly associated with transverse hemitropy \eqref{eq:defhemitropy}. So, for hemitropic rods (and even isotropic) one may use the same helical growth law \eqref{eq:pitchfix} without any notion of microstructural pitch, in which case both $\gamma$ and $\mathcal{M}$ are independent growth parameters. For such a growth law all helices with pitch $\mathcal{M}$ remain unaltered under the growth map, so we denote it as the \textit{characteristic pitch} of growth.

\subsubsection{Calculation of Strains} 
The grown configuration is obtained by imposing environmental and boundary effects on the virtual stress-free configuration. Hence the strain energy is a function of $\textcolor{black}{\displaystyle\frac{\partial\widehat{\r}}{\partial s} }$, $\widehat{\Rt}$ and $\textcolor{black}{\displaystyle\frac{\partial\widehat{\Rt}}{\partial s} }$. We define the vector fields
\begin{align}
\widehat{\boldsymbol\nu}= \textcolor{black}{\frac{\partial\widehat{\r}}{\partial s} }
\quad\text{and}\quad
\widehat{\k}= \text{axial}\textcolor{black}{\bigg(\frac{\partial\widehat{\Rt}}{\partial s} } \widehat{\Rt}^T \textcolor{black}{\bigg)}.
\end{align}
Let their components be $\widehat{\boldsymbol\nu}=\widehat{\nu}_i \di$ and $\widehat{\k}=\widehat{\kappa}_i \di$ with respect to the director frame in current configuration. 
Consider the derivative
% Now let us look at the derivatives of the current field of directors with respect to the reference parameter.
\begin{subequations}
\begin{align}
\frac{\partial \di}{\partial S}=\frac{\partial \Q}{\partial S}\Q^{-1} \di 
&=\bigg[\frac{\partial \Rt}{\partial S}\W+\Rt\frac{\partial \W}{\partial S}\bigg]\W^{-1}\Rt^{-1} \di \\
&=\Big(\frac{\partial \Rt}{\partial s}\Big)\Big(\frac{\partial s}{\partial S}\Big)\Rt^{-1} \di+\Rt\frac{\partial \W}{\partial S}\W^{-1} \widetilde{\mathbf{e}}_i\\
&=\gamma\widehat{\k}\times \di+\Rt(\w\times\widetilde{\mathbf{e}}_i )\\
&=(\gamma\widehat{\k}+\Rt\w)\times \di .
\end{align}
\end{subequations}
Now define the axial vector $\b:=\text{axial}\bigg(\displaystyle\frac{\partial \Q}{\partial S}\Q^{-1}\bigg)$ which along with the straight growth assumption implies 
\begin{align}
\b=\gamma\widehat{\k}+\omega\d3. \label{eq:appcurvtostrain}
\end{align}
Given the growth parameters, this relation will be used in retracting the actual strains from the apparent curvature $\b$.
Corresponding to $\widehat{\boldsymbol\nu}$ and $\widehat{\k}$ we define 
\begin{align}
\boldsymbol\nu= \r\,' 
\quad\text{and}\quad
\k= \text{axial}\big(\Rt'\,\Rt^T \big) ,
\end{align}
along with their convected components $\boldsymbol\nu=\nu_i \di$ and $\k=\kappa_i \di$. These speeds and curvatures can be related to the actual strains by
\begin{align}
\nu_i=\gamma \,\widehat{\nu}_i 
\quad\text{and}\quad
\kappa_i=\gamma\, \widehat{\kappa}_i . \label{eq:gammafactor}
\end{align}
\textcolor{black}{With the unshearability constraint in place we have $\nu_\alpha=\widehat{\nu}_\alpha=0$, while $\nu_3=\gamma\widehat{\nu}_3$ represents the multiplicative decomposition for length-wise growth.}
\textcolor{black}{Using} the energy density function \eqref{densnearquad}, 
% The energy density function assumed is (\ref{densnearquad}) where the stains to be taken are $\widehat{\nu}_i$ and $\widehat{\kappa}_i$, so that 
internal force $\n(S)=n_i(S) \di(S)$ and moment $\m(S)=m_i(S) \di(S)$ in the current configuration can be related to the strains as follows:
\begin{align}
n_{3}&=g(\widehat{\nu}_3 )+A\widehat{\kappa}_{3},    \label{eq:const1}\\
m_{3}&=A(\widehat{\nu}_3 -1)+B\widehat{\kappa}_{3}, \label{eq:const2}  \\
m_\alpha&=C\,\widehat{\kappa}_\alpha .\label{eq:const3}
\end{align} 
\subsubsection{Equilibrium equations} 
The local linear and angular momentum balance equations for static equilibrium \citep{o2013growth,goriely2017mathematics} are as follows:
\begin{align}
\frac{\partial\n}{\partial s}+\mathbf{f}=\mathbf{0} ,\\
\frac{\partial\m}{\partial s}+\frac{\partial\r}{\partial s}\times\n+\mathbf{l} =\mathbf{0} .
\end{align}
where $\mathbf{f}$ and $\mathbf{l}$ respectively denote the body force and body moment per unit virtual arc-length. \textcolor{black}{The change of variable $\displaystyle\frac{\partial}{\partial s}(\cdot)=\frac{1}{\gamma}\frac{\partial}{\partial S}(\cdot)$} to reference coordinates results in 
\begin{align}
\n'+\gamma\mathbf{f}=\mathbf{0} ,\\
\m'+\r'\times\n+\gamma\mathbf{l} =\mathbf{0}.
\end{align}
\section{Growing rod with guided-guided ends} \label{sec:guidedeg}

\begin{figure}
\centering
\includegraphics[scale=0.64]{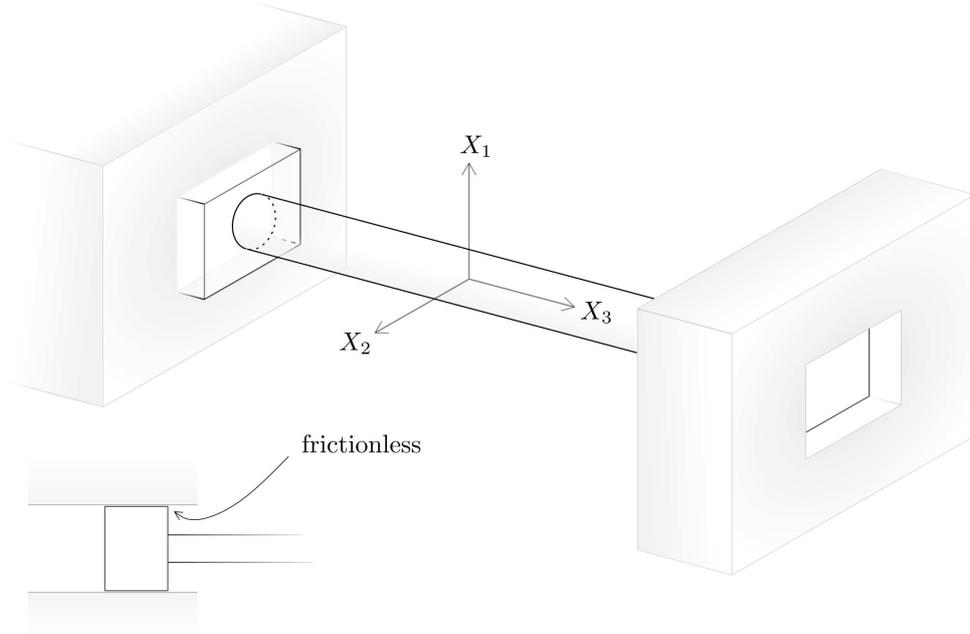}
\caption{Schematic of \textcolor{black}{g}uided-\textcolor{black}{g}uided boundary condition\textcolor{black}{: the guides arrest all degrees of freedom at the ends except for axial translation.}}
\label{fig3}
\end{figure}

% means that the rod is fixed to a block which is constrained by a slot to translate in a particular direction, any other translation or rotation of the block is arrested by the guide slot. We assume the guide slots at both ends to be coaxial with the rod itself.
%
\textcolor{black}{
%Guided-guided ends are a less restricted boundary condition pair than fixed-fixed, hence easier to solve mathematically.
%It is known that a fixed-free hemitropic rod subject to an axial load always deforms in plane, while a fixed-fixed hemitropic rod subject to axial displacements can attain non-planar buckled states. 
%The fixed-free setup subject to axial load is axially constrained by not rotationally; whereas the fixed-fixed rod subject to axial displacement or load is constrained both axially and rotationally.
A fixed-fixed rod subject to axial displacement or load is constrained both axially and rotationally, and is known to buckle out-of-plane with a transversely hemitropic consitutive law. 
But in a fixed-free hemitropic rod subject to an axial load, material chirality does not lead to any chiral deformation and the solution is always planar \citep{healey2013bifurcation}. 
%Therefore, we were interested to see if something intermediate, that is rotationally constrained, but axially free --- like the guided-guided boundary condition--- can give rise to non-planar buckling. 
For a growing rod, there can be another intermediate boundary condition pair--- with guided ends--- which is rotationally constrained, but axially free at both ends.}
A guided boundary condition is equivalent to fixing the end of the rod to a block constrained by a slot to translate only along the rod's axis
\textcolor{black}{(Figure \ref{fig3}). In this section we show that a growing rod with guided ends can give rise to non-planar chiral solutions by itself, without any additional load.}
We use the energy function  \eqref{densnearquad} and the growth law  \eqref{eq:pitchfix} to model the rod. 
Even though all the calculations would be similar, the results can be discussed separately for two different problems -- first, a hemitropic rod and second, a rod with $n$-fold helical symmetry.

 The linear and angular momentum balance equations are 
\begin{align}
\frac{d}{ds} \bigg[ n_\alpha \Q\mathbf{e_\alpha}+\big\{ g(\widehat{\nu}_3)+A\widehat{\kappa}_3\big\}\,\Q\mathbf{e_3} \bigg]&=\mathbf{0} ,\label{eq:bvpstart}\\
\frac{d}{ds} \bigg[ C\widehat{\kappa}_\alpha\Q\mathbf{e_\alpha}+\big[ A\{\widehat{\nu}_3-1\}+B\widehat{\kappa}_3\big]\Q\e3 \bigg] & \nonumber \\
+\widehat{\r}\,'\times \Big[n_\alpha\Q\mathbf{e_\alpha}+\big\{ g(\widehat{\nu}_3)+A\widehat{\kappa}_3\big\}\Q\e3 \Big]&=\mathbf{0},  \label{eq:bvpangmombal} 
\end{align}
along with the boundary conditions
\begin{align}
\n\Big(\pm\frac{1}{2}\Big)\cdot\e3 &=0 ,\label{eq:nbc}\\
 \r\Big(\pm\frac{1}{2}\Big)\cdot\mathbf{e}_\alpha &=0  \\ \text{and} \quad
  \Q\Big(\pm\frac{1}{2}\Big)&= \mathbf{I} . 
\end{align}
The unshearability constraint \eqref{eq:newunshearability} results in
% In addition, we assume the rod to be unshearable, hence the condition
\begin{align}
\mathbf{r'}\cdot\Q\ea=0.\label{eq:bvpend} 
\end{align}
Equations \eqref{eq:bvpstart}-\eqref{eq:bvpend} comprise our boundary value problem to be solved for the fields $\r$, $\Rt$, and $n_\alpha$.
%are the fields whose solutions need to be determined. 
Since we have not imposed any sort of axial constraint, with these set of boundary conditions we will get a family of solutions differing by a scalar multiple of $\e3$.

The rod is assumed to be of unit length; thus, all the kinematic quantities are dimensionless by default. 
The components of internal force, internal moment, material constants $A,B$ and the response function $g(\cdot)$ can be all non-dimensionalized against $C$ by either dividing the concerned quantities in \eqref{eq:const1}-\eqref{eq:const3} by C, or equivalently setting $C=1$ in the boundary value problem \eqref{eq:LM1,ed1}-\eqref{eq:BC3,ed1}.
We follow the bifurcation analysis methodology presented by \cite{smith2008predicting,healey2013bifurcation} wherein first a primary solution is determined  which is then perturbed and the boundary value problem is rederived in terms of the perturbations to get linearized equations.

\subsection{The \textcolor{black}{straight} solution}
Let us consider a solution where the rod always remains straight while growing, \textcolor{black}{given by}
\begin{align}
\mathbf{r}(S)=\lambda S \e3 
\quad,\quad
 \Q(S)=\mathbf{I} 
 \quad,\quad
  n_\alpha(S)=0  \,.
\end{align}
where $S\in \big[-\frac{1}{2},+\frac{1}{2}\big]$. This solution has its local force, moment and strain fields as follows:
\begin{align}
\widehat{\boldsymbol\nu}(s)&=\frac{\lambda}{\gamma}\e3,\\
\widehat{\k}(s)&=-\frac{\omega}{\gamma}\e3 ,\\
\n(S)&=\Big[g\Big(\frac{\lambda}{\gamma}\Big)-A\frac{\omega}{\gamma}\Big]\,\mathbf{e_3} ,\\
\mathbf{m}(S)&=\Big[A\Big\{\frac{\lambda}{\gamma}-1\Big\}-B\frac{\omega}{\gamma}\Big]\mathbf{e_3}.
\end{align}
For such a solution to comply with the force boundary condition \eqref{eq:nbc} we require $\lambda$ to satisfy 
\begin{align}
g\Big(\frac{\lambda}{\gamma}\Big)=A\frac{\omega}{\gamma},
\end{align}
\textcolor{black}{where $A$ denotes the degree of hemitropy and $g(\cdot)$ is the axial force response function.
As we approach a `no-growth' stage, the strain $\widehat{\nu}_3=\displaystyle\frac{\lambda}{\gamma}\rightarrow 1$ and the ratio 
\begin{align}
\frac{m_3}{\widehat{\kappa}_3}= B-\frac{A^2\big\{\frac{\lambda}{\gamma}-1\big\}}{g\big(\frac{\lambda}{\gamma}\big)}
\end{align}  
approaches $B-\displaystyle\frac{A^2}{g'(1)}$, which represents the torsional rigidity \citep{papadopoulos1999nonplanar}.}
\subsection{Perturbed Solution} \label{sec:prtbnsolv}
Consider a first order perturbation of the straight solution (with $0<\eps\ll 1$) given by
\begin{align}
\mathbf{r}(S)&=\lambda S\mathbf{e_3} + \eps \rrho(S), \label{eq:perturb1} \\
\Q(S)&=e^{\eps\mathbf{\Psi}(S)} ,\label{eq:perturb2}\\
n_{\alpha}(S)&=\eps\eta_\alpha(S) ,\label{eq:perturb3}
\end{align} 
where $\mathbf{\Psi}(S)$  is skew symmetric with $\text{axial}(\mathbf{\Psi})=:\psv$.
We require these perturbed fields to satisfy our boundary value problem. Plugging in the perturbations \eqref{eq:perturb1}-\eqref{eq:perturb3} into our boundary value problem \eqref{eq:bvpstart}-\eqref{eq:bvpend} results in the following linearized problem\textcolor{black}{:} 
\begin{align} 
\etal'\ea &=\mathbf{0},\label{eq:LM1,ed1}\\
(\psv''+\omega\e3\times\psv')\cdot\ea\,\ea +\big[A\{\lambda-\gamma\}-B\omega \big]\psv'\times\e3+\gamma\lambda\e3\times\etal\ea &= \mathbf{0} ,\label{eq:AM1,ed1}\\
\Big\{g'\Big(\frac{\lambda}{\gamma}\Big)\rrho''+A\psv''\Big\}\cdot\e3&=0 ,\label{eq:LM2,ed1}\\
(A\rrho''+B\psv'')\cdot\e3 &=0 ,\label{eq:AM2,ed1}\\
\big(\rrho'-\lambda\psv\times\e3\big)\cdot\ea &=0 ,\label{eq:unshearability,ed1}\\
\psv\Big(\pm\frac{1}{2}\Big)&=\mathbf{0},\label{eq:BC1,ed1}\\
\rrho\Big(\pm\frac{1}{2}\Big)\cdot\ea &=0 ,\label{eq:BC2,ed1}\\
\bigg[g'\Big(\frac{\lambda}{\gamma}\Big)\rrho'\Big(\pm\frac{1}{2}\Big)+A\psv'\Big(\pm\frac{1}{2}\Big)\bigg]\cdot\e3 &=0 ,\label{eq:BC3,ed1}
\end{align}
with details provided in Appendix \ref{Apx:B}.
%
% Our boundary value problem (\ref{eq:LM1,ed1})-(\ref{eq:BC3,ed1}) is very similar to that in \citep{healey2013bifurcation}. Details of linearization are present in Appendix \ref{Apx:B}.
%
Since  $Bg'\big(\frac{\lambda}{\gamma}\big)-A^2$ is non-zero (assumed to be positive), equations \eqref{eq:LM2,ed1} and \eqref{eq:AM2,ed1} imply 
\begin{align}
\rrho''\cdot\e3 = 0 \quad \text{and}\quad
\psv''\cdot\e3 = 0.
\end{align}
Boundary condition \eqref{eq:BC1,ed1} forces us to have $\psv(S) \in \text{span}\{\eon,\eto\}$, which motivates the introduction of the decomposition
\begin{align}
\rrho(S)&=\rrho_t(S)+\rrho_a(S), 
\end{align}
where $\rrho_t(S) \in \text{span}\{\eon,\eto\}$ and $\rrho_a(S) \in \text{span}\{\e3\}$.

Equations \eqref{eq:LM1,ed1}-\eqref{eq:BC3,ed1} can now be reduced to the following (details in Appendix \ref{Apx:C}):
\begin{align}
\psv''+\zeta\psv'\times\e3 &=\psv'\Big(+\frac{1}{2}\Big)-\psv'\Big(-\frac{1}{2}\Big) ,\label{eq:VecODE,ed1} \\
\rrho_t'&=\lambda\psv \times \e3 \,, \label{eq:unshearability,ed2} \\
\rrho_a'' &= \mathbf{0},
\end{align}
accompanied by the boundary conditions 
\begin{align}
\rrho_t\Big(\pm\frac{1}{2}\Big)&=\mathbf{0},\label{eq:BC2,ed2} \\
\rrho_a'\Big(\pm\frac{1}{2}\Big) &= \mathbf{0}.
\end{align}
The new parameter $\zeta$ appearing in \eqref{eq:VecODE,ed1} is defined as
\begin{align}
\zeta :=A(\lambda-\gamma)-(B+1)\omega.\label{def:parzeta}
\end{align}
It is clear that $\rrho_a(S)=C_o\e3$ for all $S$, where $C_o$ is a constant that appears because we have put no physical constraint in axial direction. As the rod can slide in the axially without causing any strain, we can fix $C_o=0$. % without any hesitation.

For $\zeta =0$, the problem admits only trivial solutions (Appendix \ref{Apx:C}). 
Now assuming $\zeta \neq 0$, the differential equations \eqref{eq:VecODE,ed1} and \eqref{eq:unshearability,ed2} admit general solutions of the form
\begin{align}
 \underline{\psi}(S)
&=\frac{C_1}{\zeta}\begin{pmatrix} \text{sin}(\zeta S) \\[5pt] -\text{cos}(\zeta S)+2S\,\text{sin}\frac{\zeta}{2}  \\[5pt]
0 \end{pmatrix} +
\frac{C_2}{\zeta}\begin{pmatrix} -\text{cos}(\zeta S)-2S\,\text{sin}\frac{\zeta}{2} \\[5pt] -\text{sin}(\zeta S) \\[5pt] 0 \end{pmatrix}+\begin{pmatrix} C_3 \\[5pt] C_4 \\[5pt] 0\end{pmatrix},\label{eq:solpsi} \\
\underline{\rho_t}(S)&=C_1\frac{\lambda}{\zeta^2}\begin{pmatrix}  -\text{sin}(\zeta S)+\zeta S^2\,\text{sin}\frac{\zeta}{2}\\[5pt]   \text{cos}(\zeta S)\\[5pt]
0 \end{pmatrix} +
C_2\frac{\lambda}{\zeta^2}\begin{pmatrix}  \text{cos}(\zeta S)  \\[5pt]    \text{sin}(\zeta S)+\zeta S^2\,\text{sin}\frac{\zeta}{2}\\[5pt]0 \end{pmatrix} \nonumber \\
& +\lambda\begin{pmatrix} C_5+C_4 S \\[5pt] C_6 - C_3 S\\[5pt] 0\end{pmatrix} ,\label{eq:solrho}
\end{align}
where $C_1\,$, $C_2\,$, $\cdots$, $C_6$ are generic integration constants in $\mathbb{R}$. The representations $\underline{\psi}$ and $\underline{\rho_t}$ are with respect to the fixed basis.
The boundary conditions \eqref{eq:BC1,ed1} when invoked into the solution \eqref{eq:solpsi} leads to
\begin{align}
(C_1-C_2)\,\text{sin}\frac{\zeta}{2}=0, \label{eq:makecases}
\end{align}
simultaneously giving
\begin{align}
C_3=  \frac{C_2}{\zeta}\text{cos} \frac{\zeta}{2}    
\quad , \quad 
C_4 =  \frac{C_1}{\zeta}\text{cos} \frac{\zeta}{2}.
\end{align}
The values of $\zeta\neq 0$ for which $\text{sin}\frac{\zeta}{2}=0$ eventually lead to the trivial solution (Appendix \ref{Apx:C}).
Therefore, we assume $C_1=C_2$, which when plugged into the general solution \eqref{eq:solrho} and forced to satisfy \eqref{eq:BC2,ed2}, leads to the condition
\begin{align}
\frac{1}{\zeta}\text{sin} \frac{\zeta}{2}-\frac{1}{2}\text{cos} \frac{\zeta}{2}=0 \label{eq:bifmode}.
\end{align}
It simultaneously leads to the constants
\begin{align}
C_5=-\frac{C_1}{\zeta^2}\Big(\frac{\zeta}{4}\text{sin} \frac{\zeta}{2}+\text{cos} \frac{\zeta}{2}\Big)=C_6.
\end{align}
 
Hence we have an out-of-plane solution,
\begin{align} 
\underline{\rho_t}(S) &=C_1\frac{\lambda}{\zeta^2}\bigg\{\text{cos}(\zeta S)+\Big(S^2-\frac{1}{4}\Big)\zeta\text{sin}\frac{\zeta}{2}-\text{cos}\frac{\zeta}{2}\bigg\}\begin{pmatrix}  1\\[5pt] 1\\[5pt]
0 \end{pmatrix} \nonumber \\
&+ C_1\frac{\lambda}{\zeta^2}\Big\{S\zeta\text{cos}\frac{\zeta}{2}-\text{sin}(\zeta S)\Big\}\begin{pmatrix}  1\\[5pt]   -1\\[5pt]
0 \end{pmatrix},    
\label{eq:solrho,case2}
\end{align}
whose existence is subject to the condition that parameters $\gamma$ and $\lambda$ admit sensible solutions ($\gamma>0$ and $\lambda>0$). %\textcolor{red}
{A positive increasing sequence $ ( a_n ) _{n=1}^\infty$ satisfying $\text{tan}(a_n)=a_n$ can be defined.}
The values taken by $\zeta \in \{\pm2 a_n : n\in\mathbb{N}\}$ correspond to the discrete bifurcation modes.

\subsection{Results and Discussion} \label{section:Discussion}
In view of the equivariance properties of our problem (Appendix \ref{Apx:D}), any  rotation of \eqref{eq:solrho,case2} about $\e3$ is an acceptable solution, hence the solution can be simplified to 
\begin{align} 
\underline{\rho}(S)&=C_1\frac{\lambda}{\zeta^2}\begin{pmatrix}  \text{cos}(\zeta S)+\big(S^2-\frac{1}{4}\big)\zeta\text{sin}\frac{\zeta}{2}-\text{cos}\frac{\zeta}{2}\\[5pt]   \text{sin}(\zeta S)-S\zeta\text{cos}\frac{\zeta}{2}\\[5pt]
0 \end{pmatrix} ,\label{sol:outplanerho}\\
\underline{\psi}(S)&=C_1\frac{\lambda}{\zeta}\begin{pmatrix}  \text{cos}\frac{\zeta}{2}-\text{cos}(\zeta S)\\[5pt]   2S\,\text{sin}\frac{\zeta}{2}-\text{sin}(\zeta S)\\[5pt]
0 \end{pmatrix} ,\label{sol:outplanepsi}\\
\eta_1(S) &= 0 ,\\
\eta_2(S) &= C_1\frac{\zeta}{\gamma}\text{cos}\frac{\zeta}{2} ,
\end{align}
where representations \eqref{sol:outplanerho} and \eqref{sol:outplanepsi} are with respect to the fixed basis.
This solution is clearly flip symmetric about $\eon$, thus making it clear that \eqref{eq:solrho,case2} was also flip symmetric, but about an axis different from $\eon$. 
The deformed centre-line $\r(S)$ for this solution is 
\begin{align}
\underline{\text{r}}(S)=\frac{\lambda}{\zeta^2}\begin{pmatrix}  \text{cos}(\zeta S)+\big(S^2-\frac{1}{4}\big)\zeta\text{sin}\frac{\zeta}{2}-\text{cos}\frac{\zeta}{2}\\[5pt]   \text{sin}(\zeta S)-S\zeta\text{cos}\frac{\zeta}{2}\\[5pt]
\zeta^2 S\end{pmatrix} ,\label{eq:fullsol}
\end{align}
represented with respect to the fixed basis, wherein $\eps C_1=1$ is set for the sake of simplicity. For a particular $\zeta \in \{\pm2 a_n : n\in\mathbb{N}\}$, the end-to-end distance $\lambda$ and growth stage $\gamma$ can be found by solving the system
\begin{align}
g\Big(\frac{\lambda}{\gamma}\Big)&=\frac{A}{\mathcal{M}}\bigg(1-\frac{1}{\gamma}\bigg) ,\label{eq:fullsolcond1}\\
\zeta&=A(\lambda-\gamma)-\frac{B+1}{\mathcal{M}}(\gamma-1) ,\label{eq:fullsolcond2}
\end{align} 
simultaneously (Table \ref{tab1}, Appendix \ref{Apx:E}). Equations \eqref{eq:fullsolcond1}-\eqref{eq:fullsolcond2} couple the axial force response of the rod with the bifurcation mode caused due to growth, via the kinematic constraint of symmetry preserving growth \eqref{eq:pitchfix}. Whenever this system does not admit a solution $\gamma>0$ and $\lambda>0$, the perturbation chosen gives only trivial solutions, indicating that out-of-plane buckling is not guaranteed. 
\textcolor{black}{Note that, linear stability analysis reveals only the shape of the buckled state, without any information on its amplitude. In the ensuing discussion, all results are described in terms of $\gamma$, $\lambda$ and $\zeta$ --- these are the fundamental geometric quantities that help us understand the interplay of growth and material chirality.}

% \vspace{10pt}
\begin{figure}
\centering
\qquad\qquad\qquad\qquad\qquad
\includegraphics[scale=1]{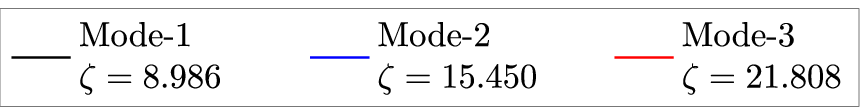} \\
\subfigure[\textcolor{black}{Spatial Deformation}]{
 \includegraphics[scale=0.465]{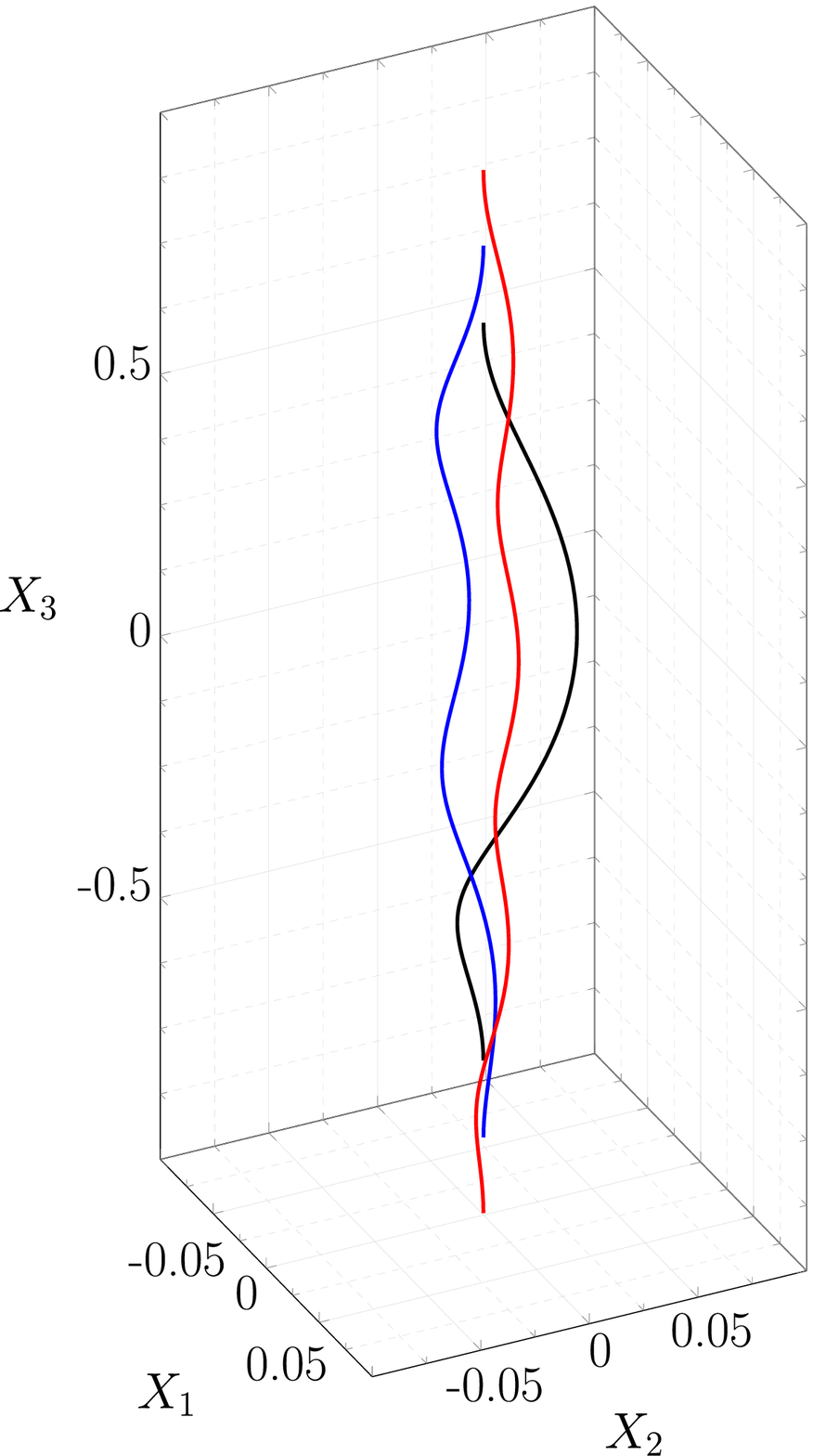}     \label{fig:subfig1}   }
\subfigure[\textcolor{black}{$X_1-X_3$ projection} ]{
 \includegraphics[scale=0.46]{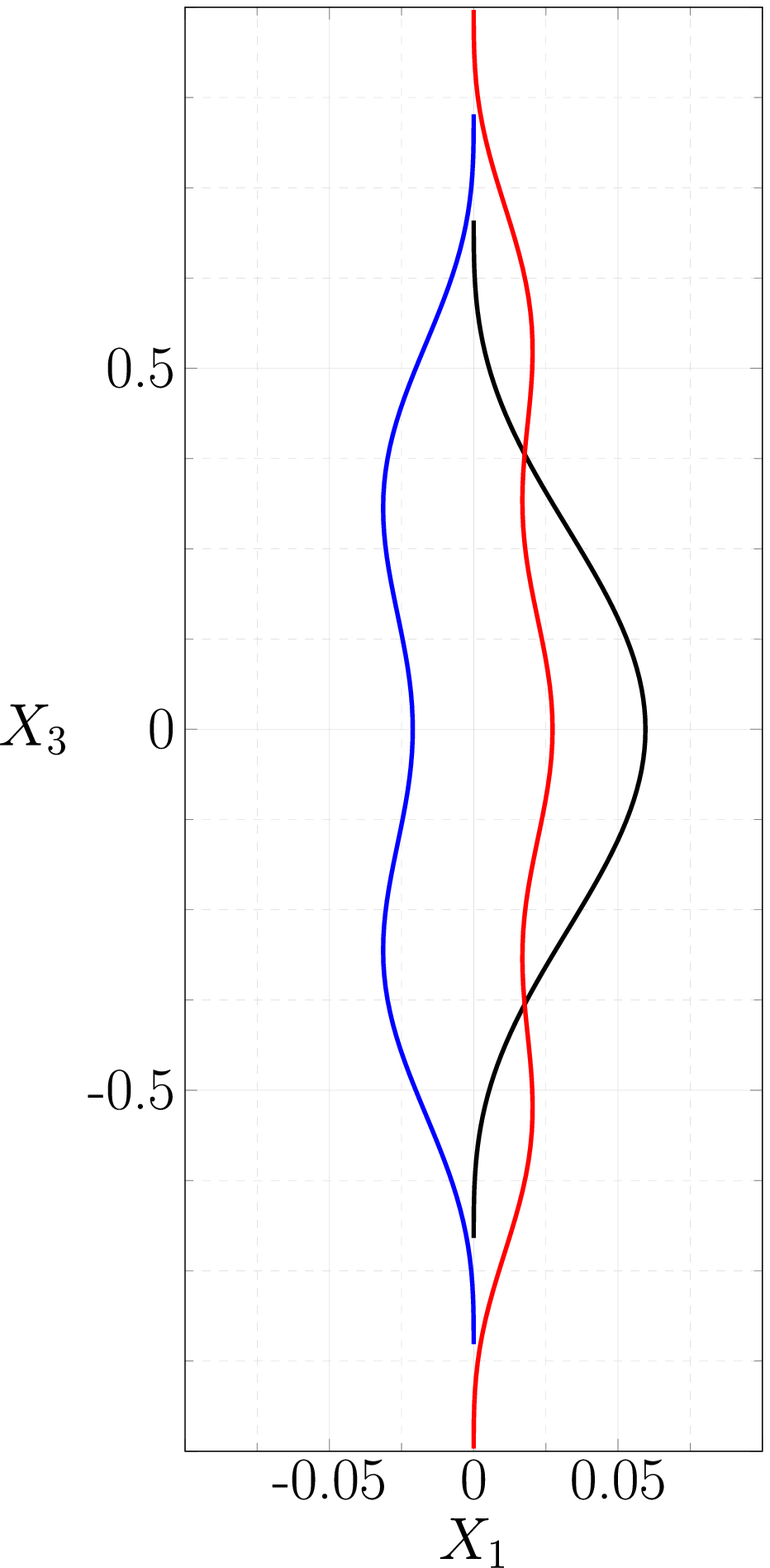}     \label{fig:subfig2}   }
\subfigure[\textcolor{black}{$X_2-X_3$ projection} ]{
 \includegraphics[scale=0.46]{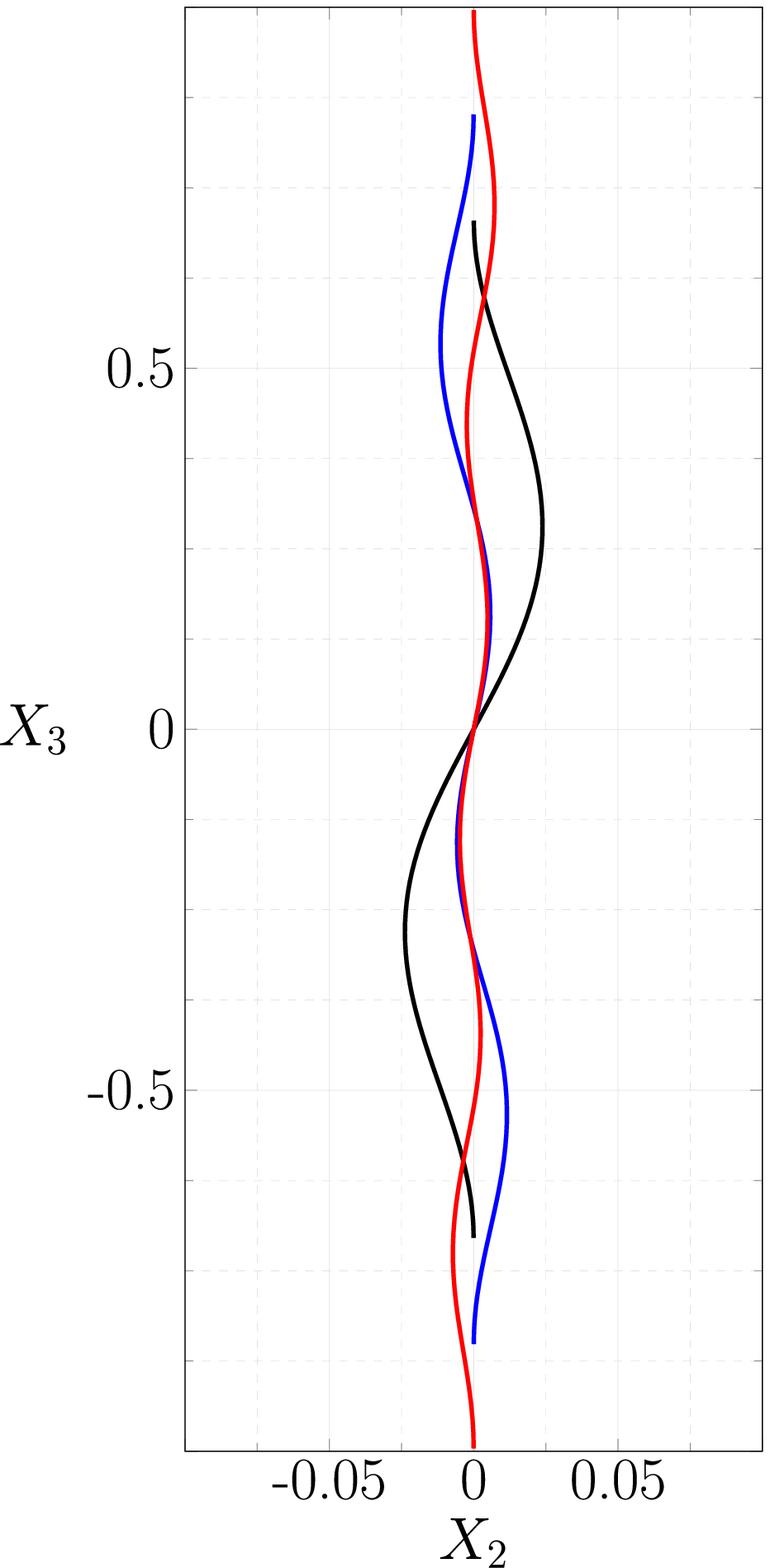}     \label{fig:subfig3}   }
 \subfigure[\textcolor{black}{$X_1-X_2$ projection}]{
 \includegraphics[scale=0.465]{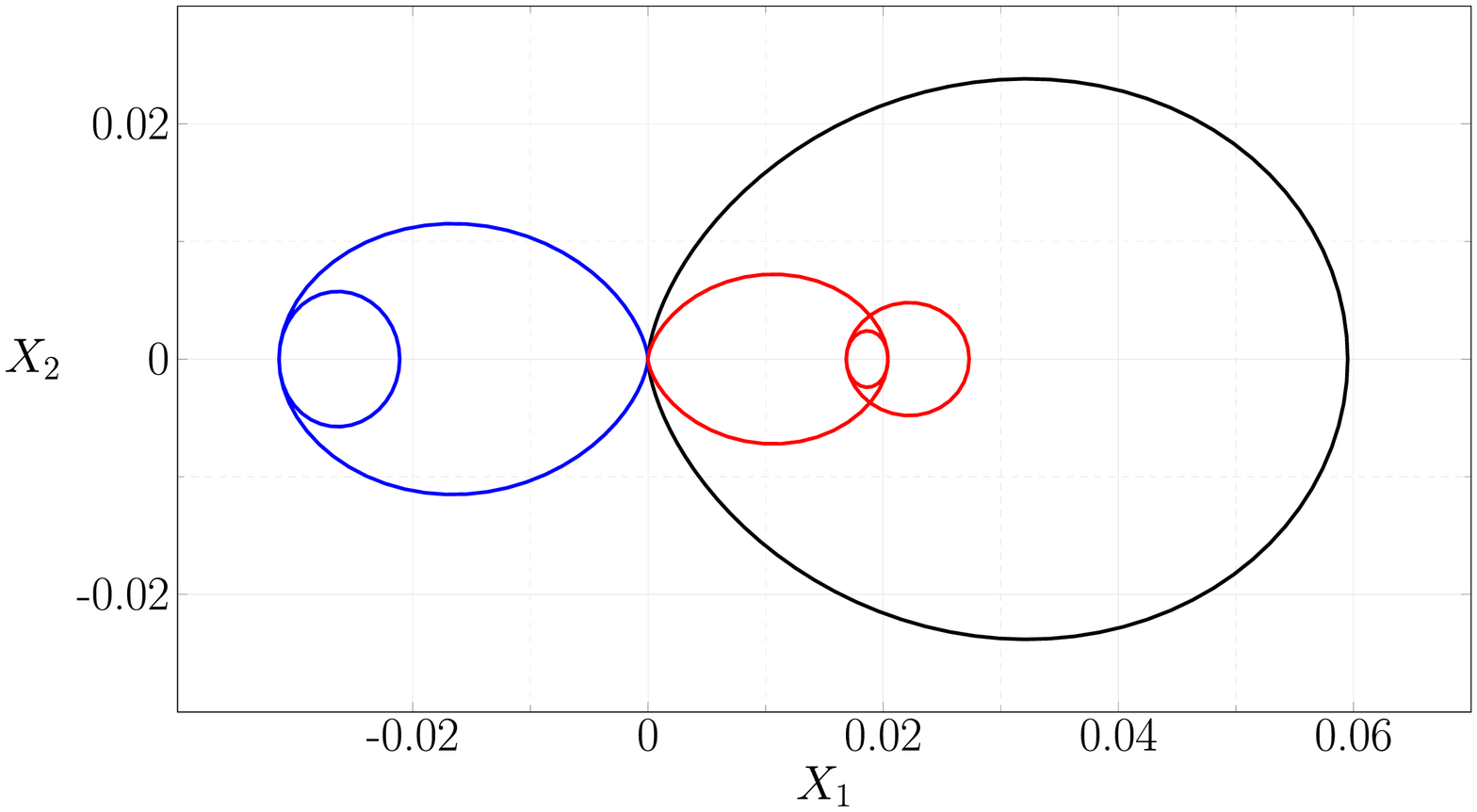}     \label{fig:subfig4}   }
\includegraphics[scale=1]{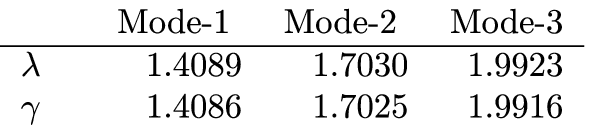}
\caption{Out-of-plane bifurcated solution for the case $\mathcal{M}=-0.1$, $A=-8$, $B=1.2$ and $F=10^5$. 
\textcolor{black}{%Spatial deformation of the rod centre-line has been shown along with its projections on the coordinate planes.
The spatial curves shown correspond to the shape of rod's centre line in the first three buckled modes. These solutions are flip-symmetric about $\eon$ (Appendix \ref{Apx:D}). The values of end-to-end distance $\lambda$ and growth stage $\gamma$ at buckling are calculated using \eqref{eq:fullsolcond1}-\eqref{eq:fullsolcond2}. Magnitude of $\zeta$ denotes the mode of bifurcated solution, while its sign controls the chirality. Reflections about $\eon-\e3$ plane correspond to the reversal in chirality $\zeta\mapsto-\zeta$.}}
\label{fig4}
\end{figure}
% \vspace{5pt}

An inspection of \eqref{eq:fullsol} reveals that the sign change $\zeta \mapsto -\zeta  $ reverses the chirality of solution curve, reflecting it about $\eon-\e3$ plane (Figure \ref{fig4}). Moreover, since our solution is flip symmetric, this is equivalent to the reflection in $\eon-\eto$ plane. \textcolor{black}{Reflections in $\eto-\e3$ plane also give solutions of opposite chirality, but need to be rotated by $180^\circ$ about $\e3$-axis to coincide with the reflections in $\eon-\e3$ plane}. These centre-line solutions with handedness are similar to those obtained by \cite{healey2013bifurcation} for a fixed-fixed rod under axial compression.

Internal chirality of the rod is taken care of by the constants $\mathcal{M}$ and $A$. In case of hemitropic rods, $A$ captures chirality in load response of the rod while $\mathcal{M}$ contains information regarding the chiral growth law. For rods with $n$-fold helical symmetry, $A$ denotes the same thing, but with the assumption of symmetry preserving growth in place, $\mathcal{M}$ captures chirality in microstructure. 

Consider two rods with opposite internal chirality with  all other material properties as same. 
Let one of them with chiral constants $A$, $\mathcal{M}$ have a solution with bifurcation mode $\zeta$, end-to-end distance $\lambda$ and growth stage $\gamma$. Naturally the second rod with opposite internal chirality is expected to give rise to a reflected solution with bifurcation mode $-\zeta$ while end-to-end distance and growth stage are still the same. Thus equations \eqref{eq:fullsolcond1}-\eqref{eq:fullsolcond2} imply that the chiral constants associated with the second rod are $-A$ and $-\mathcal{M}$. We infer that the complete reversal of internal chirality in rods requires the transformations $\mathcal{M}\mapsto -\mathcal{M}$ and $A\mapsto -A$ to be taken simultaneously. In addition, the $\zeta$ solution of a rod with internal chirality $\mathcal{M},A$ and the $-\zeta$ solution of a rod with opposite internal chirality $-\mathcal{M},-A$ are mirror images with respect to $\eon-\eto$ and $\eon-\e3$ planes.

\textcolor{black}{
%The growth stage $\gamma$ is an important measure as it helps us to see the stage of growth when the rod buckles out-of-plane. Had the guides been absent, this $\gamma$ would have been the end-to-end distance of the rod at that stage of growth. It is natural to ask how this end-to-end distance gets influenced by the presence of guides, which is why we compare $\lambda$ and $\gamma$.
%Moreover, linear stability analysis gives us only the end-to-end distance and the shape of the non-planar solution, without any information on the amplitude of the solution. So to understand the interplay of chirality in growth and constitutive law, we only have $\lambda$ and $\zeta$ available to us. 
In the absence of guides, the end-to-end distance of the rod would have been the same as its growth stage $\gamma$. Therefore, in order to understand the influence of guides on this end-to-end distance, we compare the growth stage $\gamma$ at which the rod buckles with the corresponding value of $\lambda$.}
Assuming that \eqref{eq:fullsolcond1}-\eqref{eq:fullsolcond2} admit an acceptable solution, the monotonicity of $g(\cdot)$ and the condition $g(1)=0$ reveal the following observations:
\begin{description}
\item [ \hspace{1pt} Growth] $\gamma>1$
\begin{itemize}
\item   $A$ and $\mathcal{M}$ are of same sign if and only if $\lambda>\gamma$, signifying that the ends in current configuration have moved away from each other, as compared to both initial and virtual configurations. 

\item   $A$ and $\mathcal{M}$ are of opposite if and only if $\lambda<\gamma$, signifying that the ends in current configuration have come closer as compared to virtual configuration, but no guaranteed comparison can be made with the initial configuration. 
\end{itemize}

\item [ \hspace{1pt} Atrophy] $\gamma<1$
\begin{itemize}
\item   $A$ and $\mathcal{M}$ are of opposite sign if and only if $\lambda>\gamma$,  signifying that the ends in current configuration have moved apart as compared to virtual configuration, but no guaranteed comparison can be made with the initial configuration.

\item   $A$ and $\mathcal{M}$ are of same if and only if $\lambda<\gamma$, signifying that the ends in current configuration have come closer as compared to both initial and virtual configurations. 
\end{itemize}
\end{description}
For a rod with $n$-fold helical symmetry with growth law assumed to be symmetry preserving, these results reveal an interesting interplay between chiralities in microstructure and load response of the rod. But for a hemitropic rod, the growth law allowing cross-section to rotate makes the guided-guided problem similar to a non-growing rod subject to a axial twist at one end while the other end is free to move axially. And the results above directly reflect the \textit{twist-extension type Poisson effect} expected in hemitropic rods.

% \vspace{10pt}
\begin{figure}
\centering
\includegraphics[scale=1]{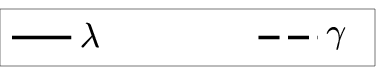} \\
\subfigure[\textcolor{black}{Rods with opposite chirality $A=-8$ and $+8$ buckle at approximately the same growth stage but their corresponding end-to-end distances satisfy $\lambda^-> \lambda^o>\lambda^+.$}]{
 \includegraphics[scale=0.64]{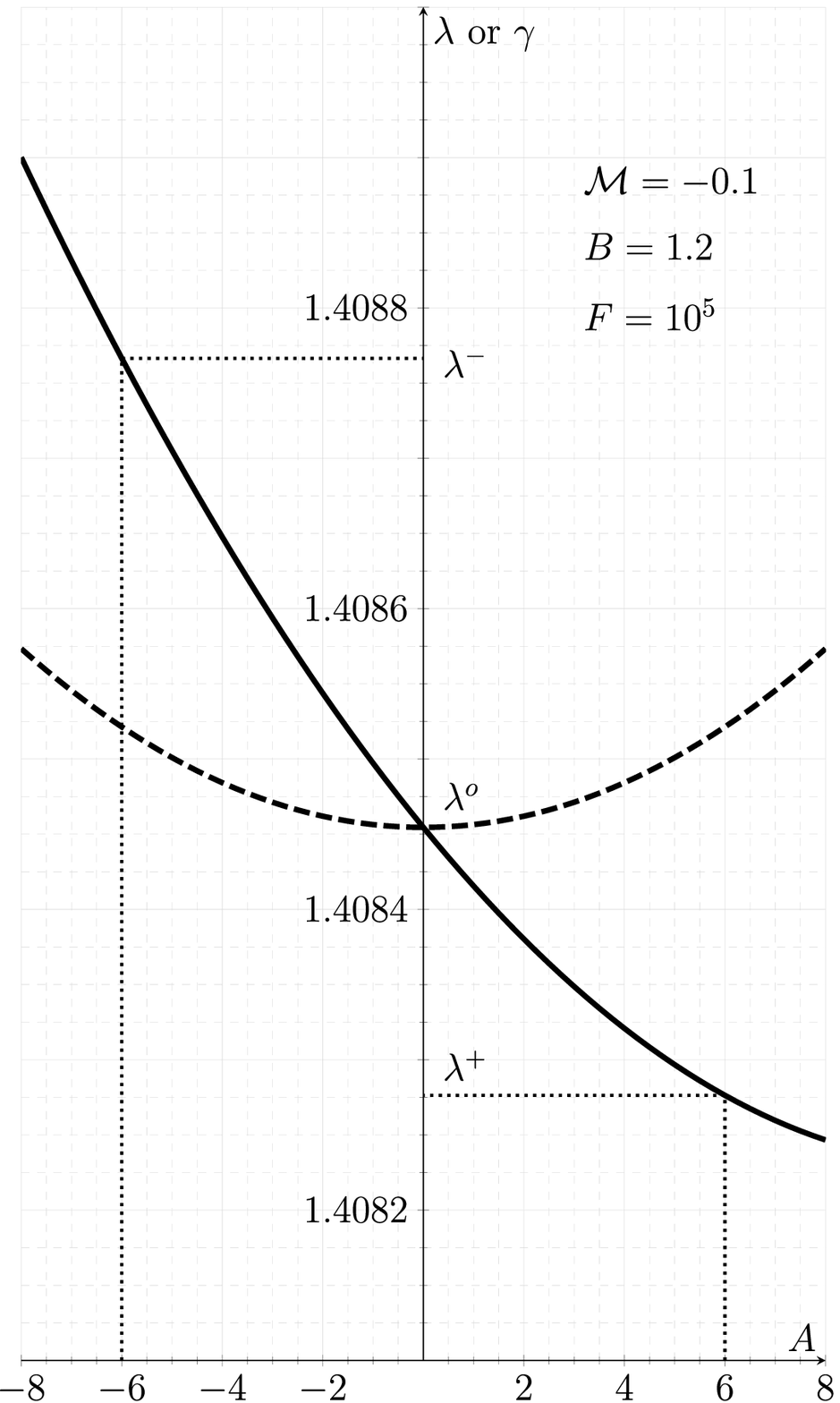}     \label{fig7:subfig1}   }
 \quad\;
 \subfigure[\textcolor{black}{Rods with opposite chirality $A=-20$ and $+20$ buckle at approximately the same growth stage but their corresponding end-to-end distances satisfy $\lambda^-> \lambda^+> \lambda^o$.}]{
 \includegraphics[scale=0.64]{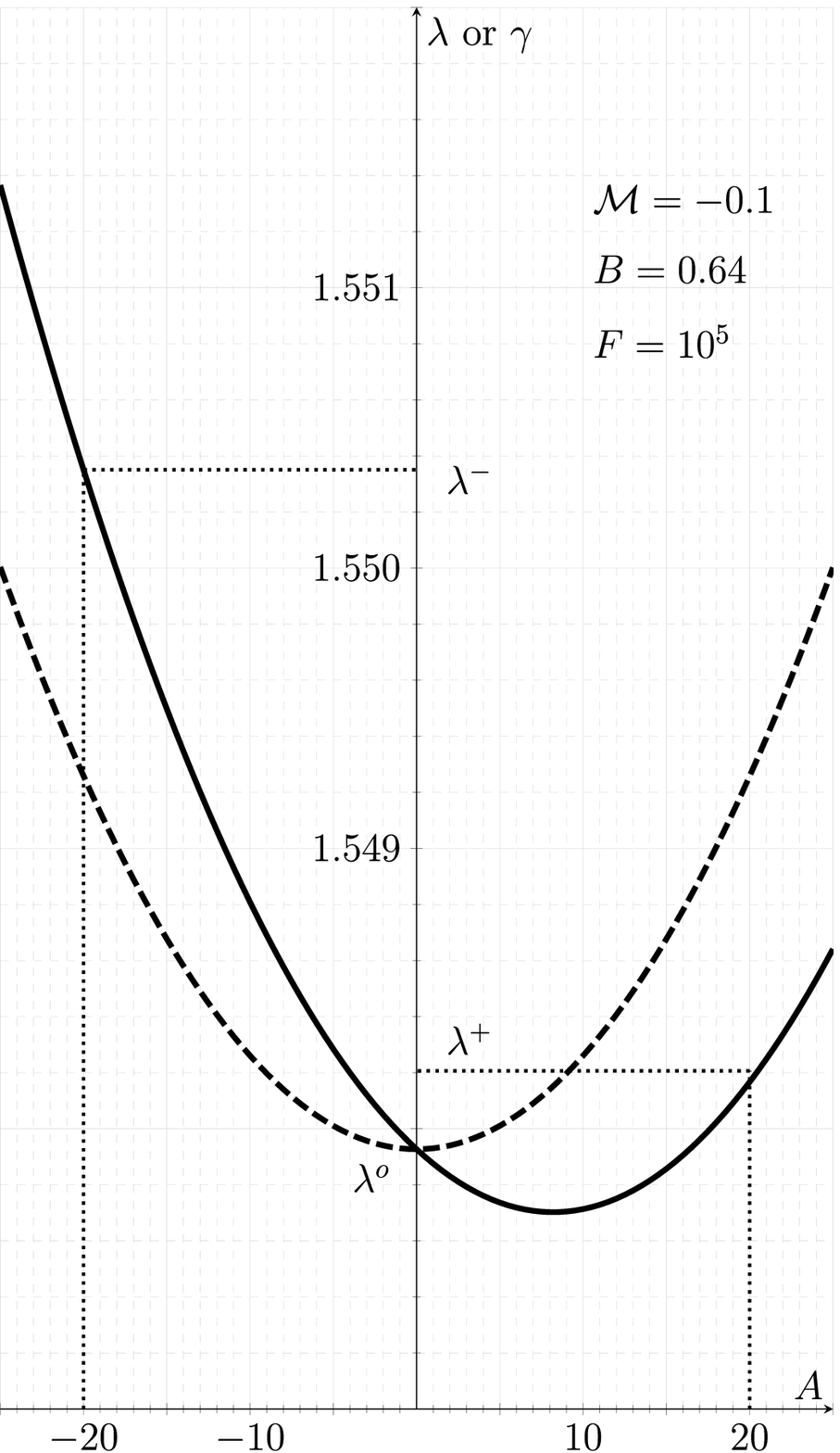}     \label{fig7:subfig2}   }
\caption{Variation of \textcolor{black}{end-to-end distance} $\lambda$ and \textcolor{black}{growth stage} $\gamma$ \textcolor{black}{at buckling} with \textcolor{black}{the degree of hemitropy} $A$ for first mode \textcolor{black}{solutions} ($\zeta=8.986$).}
\label{fig5}
\end{figure}
% \vspace{5pt}

\subsubsection*{Case of Isotropy $A=0$}

In this case, the solutions have $n_3=0$ with
\begin{align}
\gamma=\lambda=1-\frac{\zeta\mathcal{M}}{B+1}.\label{isosol}
\end{align}
A growing isotropic rod has an out-of-plane solution with sign of $\zeta$ opposite to that of $\mathcal{M}$. 
But for a decaying isotropic rod, \eqref{isosol} guarantees an out-of-plane solution only if $|\mathcal{M}|<\displaystyle\frac{B+1}{2a_1}$, and hence such solutions exist only up to the first few modes (for the chosen perturbation), with sign of $\zeta$ same as that of $\mathcal{M}$.

For small $A\neq 0$, the solution is close (in terms of $\gamma$ and $\lambda$) to that of the isotropic case with $B$, $g(\cdot)$, $\mathcal{M}$ and $\zeta$ kept same. In addition, the chirality of these solutions are same as that of the corresponding isotropic case. 
With $A\neq 0$, growing rods admit $\zeta\mathcal{M}>0$ and atrophying rods admit $\zeta\mathcal{M}<0$ only if A is taken to be very large, which in turn may be unrealistic.

Consider two rods with degrees of hemitropy $A^+>0$ and $A^-<0$, such that $A^+ +A^-=0$,  everything else being kept same. Then one of these cases gives a solution where ends come closer, while the ends move apart in the other case (comparisons made here are with respect to the virtual configuration).  Let $\lambda^+$ and $\lambda^-$ denote the respective solutions for $A^+$ and $A^-$, whereas $\lambda^o$ denotes the same for the isotropic case. While $\lambda^o$ may lie between $\lambda^+$ and $\lambda^-$, it is also a possibility that  both $\lambda^+$ and $\lambda^-$ might lie on the same side of $\lambda^o$ (Figure \ref{fig5}), thus suggesting that no definitive comment can be made on this.

\section{Conclusion}
\label{sec: conclusions}

In this work we study the growth of slender elastic rods with chiral material symmetries-- transverse hemitropy and multi-fold dihedral helical symmetry. 
Based on the intuitive notion that rods with helical symmetry should twist during growth, we propose a homogeneous growth law that allows for relative rotation of cross-sections.  
A guided-guided rod set-up is considered to illustrate the occurrence of out-of-plane buckling at certain stages of growth (or atrophy). These solutions obtained are flip symmetric and chiral in nature. 
A complete mirroring of the rod, including both growth and constitutive properties gives a solution with opposite chirality, under the same deformation. 
We show that the end-to-end distance at bifurcation modes for  the isotropic case need not lie between those for rods of opposite material chiralities, with rest of the elastic and growth properties kept same.
End-to-end distance for different combinations of growth (atrophy) and material chiralities have also been examined to understand the effect of twisting growth on the constitutive twist-extension coupling. 

Embedding our biologically active (growth or atrophy) chiral rod set-up in an elastomeric matrix and introducing inhomogeneities similar to \citep{almet2019post}, can be an interesting direction to explore. One can also consider a ply of biologically active rods, like growing bi-rods in \citep{lessinnes2017morphoelastic}, to study the effect of growth and material chiralities of individual rods on the total deformation.

\newpage

\addcontentsline{toc}{section}{Appendix}
\appendix

\section{The Growth Map} \label{Apx:A}
\textcolor{black}{Chunks of calculation skipped in Section \ref{sec:homgrowthkine} are detailed in this appendix.}
\subsection*{$\mathbf{\Lambda}$ and $\mathbf{\Omega}$ are skew-symmetric}
Since $\W(S)\in SO(3)$, \textcolor{black}{we differentiate the condition of orthogonality with respect to $S$, to obtain } 
\begin{subequations}
\begin{align}
\W\W^T=&\mathbf{I}=\W^T\W \\ 
\implies \frac{\partial\W}{\partial S}\W^T+\W\frac{\partial(\W^T)}{\partial S}=&\mathbf{O}=\frac{\partial(\W^T)}{\partial S}\W+\W^T\frac{\partial \W}{\partial S} \\
\implies 
\underbrace{\frac{\partial\W}{\partial S}\W^T}_{\mathbf{=\Omega}}
+\underbrace{\W\bigg(\frac{\partial\W}{\partial S}\bigg)^T}_{\mathbf{=\Omega}^T}
=&\mathbf{O}
=\underbrace{\bigg(\frac{\partial \W}{\partial S}\bigg)^T\W}_{=\mathbf{\Lambda}^T}
+\underbrace{\W^T\frac{\partial \W}{\partial S} }_{=\mathbf{\Lambda}}.
\end{align}
\end{subequations}
Thus $\mathbf{\Lambda}^T=-\mathbf{\Lambda}$ and $\mathbf{\Omega}^T=-\mathbf{\Omega}$;  \textcolor{black}{that is, $\mathbf{\Lambda}$ and $\mathbf{\Omega}$ are skew symmetric.}
\subsection*{Relation between $\a$ and $\w$}
We observe that $\W$ being a proper rotation must satisfy $\W=\W^*$, where $\W^*$ denotes the cofactor of $\W$. Now for any $\v\in \mathbb{E}^3$,
\begin{subequations}
\begin{align}
\w\times\v=\mathbf{\Omega}\v&=\W\mathbf{\Lambda}\W^T \v\\
&=\W(\a\times\W^T \v)\\
&=(\W^*\a)\times(\W^*\W^T \v)=\W\a \times \v.
\end{align}
\end{subequations}
This implies $\w=\W\a$.

\subsection*{Solving for $\W(S)$}
\textcolor{black}{In order to solve \eqref{Wdiffeqn} for  $\W(S)$, we first define} orthogonal tensor fields $\pphi:=e^{S\mathbf{ \Lambda } }$ and $\U:=\pphi\W^{-1}$. Then we have the following:
\begin{align}
\frac{\partial \pphi}{\partial S}&=\mathbf{ \Lambda } \pphi=\pphi\mathbf{ \Lambda } , \\
\pphi^T\frac{\partial \pphi}{\partial S}&= \W^T\U^T\frac{\partial \U}{\partial S}\W +\mathbf{ \Lambda } \, .
\end{align}
thus implying that $\U(S)$ is a constant equal to $e^{S_o\mathbf{ \Lambda } }$, which results in  
\begin{align}
\W(S)=e^{(S-S_o)\mathbf{ \Lambda } }.
\end{align}

\section{ Derivation of Perturbed Equations} \label{Apx:B}
\subsection*{Perturbations}
\textcolor{black}{In this appendix, we list down the expressions for strain fields, perturbed to first order in $\eps$. These arise from the perturbed solution \eqref{eq:perturb1}-\eqref{eq:perturb3}.}

\textcolor{black}{We begin by calculating the series expansions for apparent speed $\nu_3$ and curvature $\b$.}
\begin{subequations}
\begin{align}
\nu_3=\r'\cdot\d3 
&=\big(\lambda\e3+\eps \rrho '\big)\cdot\big(e^{\eps\ps}\e3\big)\\
&=\big(\lambda\e3+\eps \rrho '\big)\cdot\big(
\e3+\eps\psv\times\e3+\cdots\big)\\
&=(\lambda+\eps\rrho'\cdot\e3+\cdots) .
\end{align}
\end{subequations}
\textcolor{black}{Now for any $\v\in \mathbb{E}^3$,}
\begin{subequations}
\begin{align}
\frac{\partial \Q}{\partial S}\Q^T\v&=\frac{\partial \Q}{\partial S}\Big(\v -\eps \psv\times\v+\cdots\Big) \\
&=\eps\psv'\times \Big( \v -\eps \psv\times\v+\cdots \Big)+\cdots \\
&= \eps \psv'\times\v+\cdots \textcolor{black}{,}
\end{align}
\textcolor{black}{which means that}
\end{subequations}
\begin{align}
\textcolor{black}{
\b= \eps \psv'+\cdots .}
\end{align}

\textcolor{black}{Equations \eqref{eq:appcurvtostrain} and \eqref{eq:gammafactor} can now be used to calculate the following expressions for the perturbations in strains. }
\begin{align}
\widehat{\nu}_3 &=\frac{1}{\gamma}(\lambda+\eps\rrho'\cdot\e3+\cdots), \label{pertstrain1}\\
\widehat{\kappa}_\alpha &=\eps\frac{1}{\gamma}(\psv'+\omega\e3\times\psv)\cdot\mathbf{e}_\alpha+\cdots \\
\text{and} \quad
\widehat{\kappa}_3 &=-\frac{\omega}{\gamma}+\eps\frac{1}{\gamma}\psv'\cdot\e3+\cdots. \label{pertstrain3}
\end{align}
\textcolor{black}{Additionally, we also calculate }
\begin{align}
\textcolor{black}{g(\widehat{\nu}_3)=g\Big(\frac{\lambda}{\gamma}\Big)+\eps\frac{1}{\gamma}g'\Big(\frac{\lambda}{\gamma}\Big)\rrho'\cdot\e3+\cdots} \\
\text{and}\quad
n_3 =g\big(\widehat{\nu}_3 \big)+A\widehat{\kappa}_3    
=\eps\frac{1}{\gamma}\bigg\{ g'\Big(\frac{\lambda}{\gamma}\Big)\rrho'+A\psv'\bigg\}\cdot\e3+\cdots\,.\label{n3perturb}
\end{align}
\subsection*{Linearization}
\textcolor{black}{To linearise the problem, we first plug in the perturbations into the  equilibrium equations and the unshearability constraint. Then we expand each term appearing in the governing equations individually, retaining only linear terms in $\eps$.} 
\subsubsection*{Linear momentum }
\textcolor{black}{Balance of linear momentum, upon substituting the perturbations \eqref{eq:perturb3} and \eqref{n3perturb}, requires the following to be equal to zero.}
\begin{align}
\frac{d\n}{dS} &=\frac{d}{dS} \bigg[ n_\alpha \Q\mathbf{e_\alpha}+\big\{ g(\widehat{\nu}_3)+A\widehat{\kappa}_3\big\}\,\Q\mathbf{e_3} \bigg]\nonumber\\
&=\eps\bigg[\etal'\ea  +\frac{1}{\gamma}\Big\{ g'\Big(\frac{\lambda}{\gamma}\Big)\rrho''+A\psv''\Big\}\cdot\e3 \e3 \bigg]+\cdots.
\end{align}
 \textcolor{black}{We equate the transverse and axial components individually to zero, thus resulting in } \eqref{eq:LM1,ed1} and \eqref{eq:LM2,ed1}. 

\subsubsection*{Angular momentum }
\textcolor{black}{First we use \eqref{pertstrain1}-\eqref{pertstrain1}, \eqref{n3perturb} and \eqref{eq:perturb3} to obtain the following simplified expansions.}
\begin{subequations}
\begin{align}
\mathbf{r}'\times\n
&=\big(\lambda\e3+\eps\rrho'\big)\times\eps\bigg[ \etal\ea  +\frac{1}{\gamma}\Big\{ g'\Big(\frac{\lambda}{\gamma}\Big)\rrho'+A\psv'\Big\}\cdot\e3 \e3 \bigg]\\       
&=\eps\lambda\e3\times\etal\ea.
\end{align}
\end{subequations}
And
\begin{align}
\frac{d\m}{dS}
&=\frac{d}{dS} \bigg[ C\widehat{\kappa}_\alpha\Q\mathbf{e_\alpha}+\Big[ A\big\{\widehat{\nu}_3-1\big\}+B\widehat{\kappa}_3\Big]\Q\e3 \bigg]\nonumber\\
&=\eps\frac{1}{\gamma}\bigg[ C(\psv''+\omega\e3\times\psv')\cdot\ea\,\ea +\big[A\{\lambda-\gamma\}-B\omega \big]\psv'\times\e3+ (A\rrho''+B\psv'')\cdot\e3\,\e3\bigg].
\end{align}
 \textcolor{black}{These are then plugged into the angular momentum balance equation \eqref{eq:bvpangmombal}}. Subsequently we equate \textcolor{black}{the transverse and axial components of the resulting equation individually to zero, giving rise to} \eqref{eq:AM1,ed1} and \eqref{eq:AM2,ed1}. 

\subsubsection*{Unshearability }
\textcolor{black}{The constraint of unshearability requires us to equate }
\begin{subequations}
\begin{align}
\r'(S)\cdot\Q(S)\ea&=\big( \lambda\e3+\eps\rrho'\big)\cdot\big( \ea+\eps\psv\times\ea+\cdots\big)\\
&=\eps\big(\rrho'-\lambda\psv\times\e3\big)\cdot\ea+\cdots
\end{align}
\end{subequations}
\textcolor{black}{to zero, thus resulting in \eqref{eq:unshearability,ed1}.}

\section{Solution for Perturbations} \label{Apx:C}
\textcolor{black}{This appendix comprises of the details missing in Section \ref{sec:prtbnsolv}. We first demostrate how the system \eqref{eq:LM1,ed1}-\eqref{eq:BC3,ed1} can be simplified to obtain \eqref{eq:VecODE,ed1}.}
Proceeding on similar lines as that of \cite{healey2013bifurcation}, we eliminate $\eta_\alpha$ to obtain a differential equation in $\psv$ alone. Integrating \eqref{eq:LM1,ed1} we get 
\begin{align}
\etal\ea =\mathbf{c},
\end{align}
for some constant $\mathbf{c}\in \text{span}\{\mathbf{e_1}, \mathbf{e_2} \}$. Having introduced the parameter $\zeta$ in \eqref{def:parzeta},
equation \eqref{eq:AM1,ed1} transforms into 
\begin{align}
\psv''+\zeta\psv'\times\e3=\gamma\lambda\mathbf{c}\times\e3,
\end{align}
which upon integration and application of boundary condition \eqref{eq:BC1,ed1} gives
\begin{align}
\gamma\lambda \mathbf{c}\times\e3=\psv'\Big(+\frac{1}{2}\Big)-\psv'\Big(-\frac{1}{2}\Big).
\end{align}
thus leading to \textcolor{black}{the differential equation} \eqref{eq:VecODE,ed1} \textcolor{black}{in $\psv$}.
\subsection*{Solution for $\boldsymbol\psi$}
\textcolor{black}{Now we explain in detail the procedure used to solve \eqref{eq:VecODE,ed1} for $\psv$.}
Denote by $\underline{\mathtt{y}}$ the two-component representation of $\psv_t'$ with respect to $\{\eon,\eto\}$ and let $\underline{\mathtt{b}}$ denote a similar representation for $\psv_t'\big(+\frac{1}{2}\big)-\psv_t'\big(-\frac{1}{2}\big)$. Define matrix $\underline{M}=\begin{pmatrix}0 & -1 \\1 & 0\end{pmatrix}$ so that \eqref{eq:VecODE,ed1} can be rewritten as  
\begin{align}
\underline{\mathtt{y}}'=\zeta\underline{M}\underline{y}+\underline{\mathtt{b}} .\label{eq:VecODE,ed2}
\end{align}
Assume $\zeta\neq 0$ for time being.
Observe that solving \eqref{eq:VecODE,ed2} is equivalent to solving 
\begin{align}
\underline{\mathtt{x}}'=\zeta\underline{M}\,\underline{\mathtt{x}} ,\label{eq:VecODE,ed3}
\end{align}
so that the general solution of \eqref{eq:VecODE,ed2} would be given by 
\begin{align}
\underline{\mathtt{y}} &=\underline{\mathtt{x}}-\frac{1}{\zeta}\underline{M}^{-1}\underline{\mathtt{b}},\\
\underline{\mathtt{b}}&=\underline{\mathtt{y}}\Big(+\frac{1}{2}\Big) - \underline{\mathtt{y}}\Big(-\frac{1}{2}\Big) =\underline{\mathtt{x}}\Big(+\frac{1}{2}\Big) - \underline{\mathtt{x}}\Big(-\frac{1}{2}\Big) .
\end{align}
Thus we have general solutions for $\underline{\mathtt{x}}$ and $\underline{\mathtt{y}}$ given by
\begin{align}
\underline{\mathtt{x}}(S)&=C_1\begin{pmatrix} \text{cos}(\zeta S) \\[5pt] \text{sin}(\zeta S)   \end{pmatrix} +
C_2\begin{pmatrix} \text{sin}(\zeta S) \\[5pt] -\text{cos}(\zeta S) \end{pmatrix} , \\
\underline{\mathtt{y}}(S)&=C_1\begin{pmatrix} \text{cos}(\zeta S) \\[6.4pt] \text{sin}(\zeta S)+\frac{2}{\zeta}\,\text{sin}\frac{\zeta}{2}   \end{pmatrix} +
C_2\begin{pmatrix} \text{sin}(\zeta S)-\frac{2}{\zeta}\,\text{sin}\frac{\zeta}{2} \\[6.4pt] -\text{cos}(\zeta S) \end{pmatrix},
\end{align}
where $C_1$ and $C_2$ are constants in $\R$.

This gives the solution for $\psv$  as \eqref{eq:solpsi}. Finally all trivial and non-trivial solutions discussed in section \ref{sec:prtbnsolv} can be summarized as follows:
\begin{description}
\item [Case-I] Assume $\zeta = 0$. Equation \eqref{eq:VecODE,ed1} with boundary condition \eqref{eq:BC2,ed1} invoked gives
\begin{align}
\psv(S)=\frac{1}{2}\Big(S^2-\frac{1}{4}\Big)\Big\{\psv'\Big(\!+\frac{1}{2}\Big)-\psv'\Big(\!-\frac{1}{2}\Big)\Big\}, 
\end{align} 
substituting which into \eqref{eq:unshearability,ed2} gives the following relation between boundary values
\begin{align}
\rrho_t\Big(\!+\frac{1}{2}\Big)-\rrho_t\Big(\!-\frac{1}{2}\Big)=-\frac{\lambda}{12}\Big\{\psv'\Big(\!+\frac{1}{2}\Big)-\psv'\Big(\!-\frac{1}{2}\Big)\Big\}\times \e3 .
\end{align}
Invoking the boundary condition \eqref{eq:BC2,ed2}, we imply
\begin{align}
\psv'\Big(\!+\frac{1}{2}\Big)=\psv'\Big(\!-\frac{1}{2}\Big),
\end{align}
thus resulting in the trivial solution $\psv(S)=\mathbf{0}=\rrho(S)$.

\item [Case-II] Assume $\zeta \neq 0$. In this case, a general solution \eqref{eq:solrho} is obtained, which subsequently gives rise to the following sub-cases based on \eqref{eq:makecases}.

\begin{itemize}
\item
Let $\text{sin}\frac{\zeta}{2}=0$ with $\zeta \neq 0$. This implies $\zeta=2n\pi$ where $n\in\mathbb{Z}\setminus \{0\}$.
Each such value of $\zeta$ gives a solution
\begin{align} 
\underline{\rho_t}(S)=\frac{C_1\lambda}{\zeta^2}\begin{pmatrix}  -\text{sin}(\zeta S)+(-1)^n \zeta S\\[5pt]   \text{cos}(\zeta S)\\[5pt]
0 \end{pmatrix} +
\frac{C_2\lambda}{\zeta^2}\begin{pmatrix}  \text{cos}(\zeta S)  \\[5pt]    \text{sin}(\zeta S)-(-1)^n \zeta S\\[5pt] 0 \end{pmatrix}+\lambda\begin{pmatrix} C_5\\[5pt] C_6 \\[5pt] 0\end{pmatrix}. \label{eq:solrho,case1}
\end{align}
But for this to agree with \eqref{eq:BC2,ed2}, we require $C_1=0=C_2$ and $C_5=0=C_6$, thus leading to a trivial solution. 

\item  Let $C_1=C_2$ with $\zeta \neq 0$.
This leads to non-trivial out-of-plane solutions \eqref{eq:solrho,case2}, which are discussed further in section \ref{section:Discussion}.

\end{itemize}

\end{description}

\section{ Equivariance Properties of Solutions} \label{Apx:D}

Let $\F$ be the tensor defined by flip action -- a $180$-degree rotation-- about $\eon$ axis and $\textcolor{black}{\Rot_\phi}$ denote the rotation tensor about $\e3$ axis as defined in \eqref{def:rot}.
\textcolor{black}{
The components of $\F$ with respect to the fixed basis can be represented as 
\begin{align}
\underline{F}=
\begin{bmatrix}
1 & 0 & 0 \\ 
0 & -1 & 0 \\ 
0 & 0 & -1
\end{bmatrix} .
\end{align} 
}
For any solution $\displaystyle\Big(\r(S)\,,\textcolor{black}{\Q}(S)\,,n_\alpha(S)\Big)$ of the boundary value problem \eqref{eq:bvpstart}-\eqref{eq:bvpend}, the  tuple 
\begin{align}
\Big(\textcolor{black}{\Rot_\phi}\r(S)\,,\textcolor{black}{\Rot_\phi}\textcolor{black}{\Q}(S)\textcolor{black}{\Rot_\phi}^T,(\textcolor{black}{\underline{\Theta}_\phi})_{\alpha\beta}n_\beta(S)\Big)
\end{align}
also solves the system \eqref{eq:bvpstart}-\eqref{eq:bvpend} for all $0 \leq  \textcolor{black}{\phi} <2\pi$ and so does 
\begin{align}
\Big(\F\r(-S)\,,\F\textcolor{black}{\Q}(-S)\F,-\underline{F}_{\alpha\beta}n_\beta(-S)\Big)\textcolor{black}{.}
\end{align}
Equivalently in terms of perturbations, any solution $\displaystyle\Big(\rrho(S)\,,\psv(S)\,,\eta_\alpha(S)\Big)$ of the boundary value problem \eqref{eq:LM1,ed1}-\eqref{eq:BC3,ed1}, generates an entire class of solutions comprising of
\begin{align}
\Big(\textcolor{black}{\Rot_\phi}\rrho(S)\,,\textcolor{black}{\Rot_\phi}\psv(S)\,,(\textcolor{black}{\underline{\Theta}_\phi})_{\alpha\beta}\eta_\beta(S)\Big)
\end{align}  for all $0 \leq  \textcolor{black}{\phi} <2\pi$ and 
\begin{align}
 \Big(\F\rrho(-S)\,,\F\psv(-S),-\underline{F}_{\alpha\beta}\eta_\beta(-S)\Big).
\end{align}    

Our boundary value problem is equivariant with respect to the action of a group generated by rotations about $\e3$ axis and flip about $\eon$ axis.

A solution is said to be flip symmetric if 
\begin{align}
\Big(\F\r(-S)\,,\F\textcolor{black}{\Q}(-S)\F,-\underline{F}_{\alpha\beta}n_\beta(-S)\Big)=\Big(\r(S)\,,\textcolor{black}{\Q}(S)\,,n_\alpha(S)\Big),
\end{align} or equivalently if the perturbations satisfy
\begin{align}
\Big(\F\rrho(-S)\,,\F\psv(-S),-\underline{F}_{\alpha\beta}\eta_\beta(-S)\Big)=\Big(\rrho(S)\,,\psv(S)\,,\eta_\alpha(S)\Big)
\end{align}
for all $S\in \big[-\frac{1}{2},+\frac{1}{2}\big]$.

These equivariance properties of solutions are explained in much greater detail in \citep{papadopoulos1999nonplanar}.

\section{ Calculation of $\lambda$ and $\gamma$} \label{Apx:E}

First of all, numerical values of $A,B,F$ and $\mathcal{M}$ are fixed. Inspired by the calibration calculations present in \citep{papadopoulos1999nonplanar}, 
for a rod of length $L=1$ with circular cross section and material constant $C=1$, radius $r$ of the cross-section can be shown to be 
\begin{align}
r=\frac{2}{\sqrt{F}},
\end{align}
where both $r$ and $F$ are dimensionless. For instance, $F=10^6$ is equivalent to consider a $1$ metre rod with diameter $4$ millimetres.
In addition, we have the following values of $\zeta$ corresponding to different bifurcation modes
\begin{align}
\zeta &\in \{\,\;\pm\, 8.986\;,\; \pm\, 15.45\;,\;\pm\, 21.808\;,\; \pm\, 28.132 \;,\;\pm\,\,34.442 \;,\,\cdots\;\}.
\end{align}

Introduce variables $x=\frac{\lambda}{\gamma}$ and $y=\frac{1}{\gamma}$. For a particular $\zeta$, equations \eqref{eq:fullsolcond1}-\eqref{eq:fullsolcond2} require us to solve 
\begin{align}
F\, \text{ln}(x)+\bigg( \frac{A^2}{B}+\frac{A^2\zeta}{\mathcal{M}\zeta-B-1}\bigg)x=\bigg( \frac{A^2}{B}+\frac{A(A+\zeta)}{\mathcal{M}\zeta-B-1}\bigg) \label{eq:finsolve}
\end{align}
 for $x$. 
Define the following solution set. 
\begin{align}
\mathcal{S}(m,c):=& \{x\,:\, \text{ln}(x)=mx+c\,,\,x\in(0,\infty) \}.
\end{align}
We observe that, 
\begin{align}
\left|\mathcal{S}(m,c)\right| =
  \begin{cases}
  1 & \text{if } m\leq 0 \\
  0 & \text{if } m>0 \text{ and } \text{ln}(m)+c+1>0 \\
  1 & \text{if } m>0 \text{ and } \text{ln}(m)+c+1=0 \\
  2 & \text{if } m>0 \text{ and } \text{ln}(m)+c+1<0 
  \end{cases}\;,
  \end{align}
where  $m,c\in \R$ and $|\cdot|$ denotes the cardinality of a set. We set 
\begin{align}
\displaystyle m=-\frac{A^2}{F}\bigg( \frac{1}{B}+\frac{1}{\mathcal{M}\zeta-B-1}\bigg) \quad\text{and}\quad   c=\frac{A}{F}\bigg( \frac{A}{B}+\frac{A+\zeta}{\mathcal{M}\zeta-B-1}\bigg).
\end{align}
Clearly $m$ is positive only when $1<\zeta\mathcal{M}<1+B$. 
Thus if $\zeta$ and $\mathcal{M}$ have opposite sign \eqref{eq:finsolve} has a guaranteed solution. Whenever they are of same sign, the choice $\displaystyle|\mathcal{M}|<\frac{1}{2a_1}$ guarantees a solution to \eqref{eq:finsolve}, although there may be several other scenarios leading to a solution.

Once we have a solution $x_o\in\mathcal{S}(m,c)$, we have corresponding
\begin{align}
y_o=\frac{\mathcal{M}A(x_o-1)-B-1}{\mathcal{M}\zeta-B-1}
\end{align}
and $\displaystyle \lambda_o=\frac{x_o}{y_o}$, $\displaystyle \gamma_o=\frac{1}{y_o}$ would give the complete solution (Table \ref{tab1}).

\begin{table}[H]
\centering
\caption{Sample calculation for $F=10^5$ and $\zeta=8.986$}
\begin{threeparttable}
  \begin{tabular}{l @{\hskip 10pt}  r @{\hskip 15pt}r@{\hskip 15pt}r @{\hskip 30pt} rr} % creating 10 columns
 % inserting double-line
    & $\mathcal{M}$ & $A$ & $B$ & $\lambda-1$ & $\gamma-1$
\\ [0.2 ex]
\hline 
Growth 
&  $-0.1$            & $-8$ & $1.2$ & $0.4089$ &  $0.4086$ \tnote{$\dagger$}  \\[0.5 ex]
&  $-0.1$            & $8$  & $1.2$ & $0.4082$ &  $0.4086$ \tnote{$\dagger$}  \\[0.5 ex]
&  $-2\times10^{-4}$ & $24$ & $0.2$ & $-2.5\times10^{-4}$ &  $1.5\times10^{-3}$ \\[2 ex]

Atrophy 
&  $10^{-4}$ & $-16$ & $0.4$  & $3.8\times10^{-4}$ & $-0.64\times10^{-3}$ \\[0.5 ex]
&  $10^{-2}$ & $-16$ & $0.32$ & $-0.0671$ & $-0.0682$ \tnote{$\ddagger$}  \\[0.5 ex] 
&  $10^{-2}$ & $16$  & $0.32$ & $-0.0693$ & $-0.0682$ \tnote{$\ddagger$}  \\
[0.2 ex]
\hline
\end{tabular}
  \begin{tablenotes}
    \item[$\dagger$,$\ddagger$] \footnotesize Values are really close.
    \end{tablenotes}
\end{threeparttable}

\label{tab1}
\end{table}

 Note that we sometimes we may get an absurd solution $y_o<0$. 
 
 For example, the case $\mathcal{M}=0.16$, $A=-8$, $B=0.4$ and $F=10^5$ when solved with $\zeta=8.986$ gives $x_o= 0.9815$, $y_o=-36.4484$, an invalid solution. Moreover, in this case we have $m=-0.0185$, indicating that there is no other valid out-of-plane deformation arising from the chosen perturbation.  
 
 \newpage
 
% \addcontentsline{toc}{section}{References}
% \renewcommand{\bibname }{References}
\bibliographystyle{author-year-prashant}
\bibliography{referencesfile}

\end{document}